\documentclass[reprint,aps,pre]{revtex4-2}

\usepackage{amsmath,amssymb,graphicx}
\usepackage{bm}
\usepackage{hyperref}
\usepackage{caption}
\usepackage{subcaption}
\usepackage{float}
\usepackage{xcolor}
\usepackage[percent]{overpic}

\begin{document}

\title{Dynamical Facilitation in Active Glass Formers: Role of Morphology and Persistence}

\author{Dipanwita Ghoshal}
\email{ghoshaldipanwita10@gmail.com}
\affiliation{Indian Institute of Technology, Madras, Chennai-600036, India\\ 
Hong Kong Baptist University, 224 Waterloo Rd, Kowloon Tong, Hong Kong
\\
Centro de Fisica de Materiales (CSIC, UPV/EHU), Paseo Manuel de Lardizabal 5, E-20018 San Sebastian, Spain}
\date{\today}

\begin{abstract}
Understanding dynamical facilitation in glass-forming systems driven far from equilibrium by active forces remains an open challenge. It remains unresolved whether this framework applies to active glasses, where persistent self-propulsion introduces directional memory and breaks detailed balance. In particular, the role of the persistent length $l_p$ as a control parameter for cooperative relaxation is understood in the present analysis in a new manner. Here, we use large-scale simulations of a two-dimensional athermal Ornstein-Uhlenbeck particle (AOUP) model to investigate how persistent active forcing modifies dynamical facilitation. We analyze the morphology of cooperatively rearranging regions (CRRs) and the spatial transport of mobility excitations. A spatially resolved core-shell decomposition reveals a contrasting sensitivity of core and shell morphology to persistence time $\tau_p$ - a core that undergoes global morphological changes and hence maintains internal plasticity, while an extended shell acts as a rigid scaffold, supporting only axial morphological change and serving as a dynamical conduit of facilitation. Activity induces a pronounced non-monotonic dependence of dynamical observables, including modal displacement, shell occupation probability, and facilitation length, on the persistence time $\tau_p$. This behavior reflects a competition between persistence time $\tau_p$ and noise strength $T_{\text{eff}}$ and generation of coherence- or trapping-dominated dynamics at larger $\tau_p$ depending on the magnitude of $T_{\text{eff}}$. Despite the morphological changes of core and shell in a CRR, the facilitation length ($\xi_{\text{fac}}$) shows an approximate diffusive-like scaling collapse when rescaled by the persistence length  $l_p=\sqrt{T_{\text{eff}}\tau_p}$. This relation is consistent with a diffusive-like time-length coupling, $\xi_{\text{fac}} \sim \tau_{\alpha}^{1/2}$, suggesting that activity modifies the geometry of facilitation pathways while preserving their large-scale transport properties. These results support a generalized facilitation picture for active glass formers, where persistence reshapes but does not eliminate the mechanisms of cooperative relaxation and its long-time transport.
\end{abstract} 
\maketitle

\section{Introduction}
\label{intro}
The glass transition remains a central challenge in condensed matter physics. It is characterized by a dramatic slowdown of structural relaxation as a liquid approaches an amorphous solid state \cite{novikov2004poisson}. A key hallmark of this phenomenon is dynamical heterogeneity, where relaxation occurs through spatially localized regions of enhanced mobility embedded within a comparatively rigid matrix \citep{berthier2005direct}. These heterogeneous dynamics originate from localized rearrangements that propagate through the system and collectively govern structural relaxation. Several theoretical frameworks have been developed to describe this phenomenon. Mode-coupling theory (MCT) describes the growth of relaxation timescales and dynamical correlations at a mean-field level \cite{das2004mode}. In contrast, dynamical facilitation (DF) theory emphasizes the role of localized mobility excitations and their spatiotemporal propagation \cite{garrahan2003coarse,chandler2010dynamics,keys2013calorimetric}. Within this framework, structural relaxation occurs through the hierarchical spreading of these localized excitations, which facilitate further motion in neighboring regions. The importance of facilitation in glassy dynamics has been increasingly emphasized in recent theoretical work \cite{costigliola2024glass}. Cooperatively rearranging regions (CRRs) represent the structural manifestation of such facilitated dynamics and encode the spatial organization of collective particle motion. In passive glass formers, DF theory models relaxation as a density-controlled process in which localized excitations propagate approximately isotropically. In this regime, facilitation lengths and relaxation timescales are primarily governed by the concentration of excitations, which in turn is controlled by temperature \cite{herrero2024direct,chacko2024dynamical,hasyim2024emergent}. Within this framework, the internal geometry of excitation clusters is typically not considered a dominant factor in determining facilitation strength. Active glass formers introduce qualitatively new ingredients into this scenario. Persistent self-propulsion generates directional memory, breaks detailed balance, and introduces an intrinsic timescale through the persistence time $\tau_p$. Unlike thermal fluctuations, active forcing is temporally correlated and can induce anisotropic stresses and coherent particle motion. These effects can significantly modify the relaxation dynamics of glassy systems. Recent theoretical approaches, including active elastoplastic models and studies of activity-induced fluidization in supercooled systems, highlight the influence of activity on glassy relaxation processes \cite{ghosh2025elastoplastic,truong2025facilitation}. A central theoretical question, therefore, arises: does facilitation in active glasses remain a density-controlled, approximately isotropic process, or does activity fundamentally reorganize the internal structure and transport pathways of excitations? While several studies have examined how activity modifies relaxation time scales and dynamical heterogeneity \cite{hasyim2024emergent,keta2022disordered,keta2024emerging}, a spatially resolved understanding of how mobility excitations are organized and transported in active systems remains incomplete. Importantly, the applicability of dynamical facilitation in such nonequilibrium systems is not guaranteed a priori. In the present work, we do not assume the validity of DF, instead we explicitly test its dynamical signatures by identifying localized, persistent, and spatially correlated mobility excitations. This allows us to assess whether facilitation provides a useful organizing principle for relaxation in active glass formers. In this work, we show that facilitation in active glasses is significantly modified by the morphology of excitation clusters. We introduce a spatially resolved core-shell decomposition of cooperatively rearranging regions (CRRs), revealing a hierarchical structure consisting of a compact core associated with localized rearrangements and an extended shell that mediates mobility transport. By analyzing shape descriptors derived from the gyration tensor, including asphericity and acylindricity, we identify persistence-controlled morphological changes that correlate with variations in facilitation strength. We further demonstrate that the facilitation length $\xi_{\mathrm{fac}}$ exhibits a pronounced non-monotonic dependence on the persistence time $\tau_p$, indicating a regime of maximal cooperative transport at intermediate persistence. This behavior reflects a competition between persistence-enhanced cage-breaking and coherence- or trapping-dominated dynamics at larger $\tau_p$, which suppresses relative motion within CRRs. At large persistence, the enhancement
of $\xi_{\text{fac}}$ correlates with directional coherence rather than increased local rearrangements, indicating a crossover in the mechanism of mobility transport. Upon rescaling by the persistence length $l_p=\sqrt{T_{\mathrm{eff}}\tau_p}$, we observe an approximate scaling collapse of the facilitation length  with the structural relaxation time, consistent with a diffusive like propagation of mobility excitations within the regime $\tau_p < \tau_\alpha$. We emphasize that this scaling should be interpreted as an effective description within the explored parameter regime rather than universal relation. Taken together, these results suggests that active glass formers remain facilitation-controlled systems, but with activity-dependent spatial organization of excitations. The key insight of this work is that persistence does not eliminate facilitation but reshapes its geometry, leading to anisotropic, morphology-dependent transport of mobility. The primary goal of the present work is not to construct a predictive theory of active glassy dynamics, but rather to provide a physically transparent dynamical framework that connects activity, morphology and facilitation. In this sense, our approaches complement existing thermodynamic and mean-field descriptions of dynamical facilitation theory by focusing on the spatial organization and transport of mobility excitations in non-equilibrium systems. The remainder of this paper is organized as follows. Section~\ref{sim_details} describes the model and simulation protocol. Section~\ref{results-1} introduces the core-shell decomposition and statistical characterization of CRRs. Section~\ref{Mobility transfer function} presents the mobility transfer analysis used to extract the facilitation length and its scaling behavior. Section~\ref{Shape Analysis} examines CRR morphology using cluster shape metrics, and Section~\ref{conclu} summarizes the implications of our findings for active glassy dynamics.
%%%%%%%%%%%%%%%%%%%%%%%%%%%%%%%%%%%%%%%%%%%%%%%%%%%%%%%%%%%%%%%%%%%%%%%%%%%%%%%%%%%%%%
\section{Model and Simulation Details}
\label{sim_details}
An athermal active Ornstein-Uhlenbeck particle (AOUP) glass former was simulated. The AOUP framework provides a minimal model of persistent active forcing without aligning interactions, isolating the role of temporal correlations in modifying glassy dynamics. The model follows the formulation presented by Ghoshal and Joy~\cite{ghoshal2020connecting}. Each particle evolves under overdamped dynamics with self-propulsion generated by an Ornstein-Uhlenbeck colored-noise process~\cite{szamel2014self}:
\begin{align}
\dot{\mathbf r}_i &= \frac{1}{m\gamma}
\left(-\sum_{j\neq i}\nabla_i \phi(r_{ij}) 
+ \mathbf f_i \right), 
\label{eq:ri} \\
\dot{\mathbf f}_i &= -\frac{1}{\tau_p}\mathbf f_i 
+ \frac{\sqrt{2 m\gamma k_B T_{\mathrm{eff}}}}{\tau_p}\,
\boldsymbol{\eta}_i(t).
\label{eq:fi}
\end{align}
Here, $m\gamma$ is the friction coefficient and $\tau_p$ is the persistence time of the self-propulsion. 
The stochastic term $\boldsymbol{\eta}_i(t)$ is Gaussian white noise with zero mean and unit variance,
\begin{equation}
\langle \eta_{i,\alpha}(t)\eta_{j,\beta}(t')\rangle
= \delta_{ij}\delta_{\alpha\beta}\delta(t-t'),
\end{equation}

which leads to exponentially correlated active forces,

\begin{equation}
\langle f_{i,\alpha}(t) f_{j,\beta}(t')\rangle
= \frac{m\gamma k_B T_{\mathrm{eff}}}{\tau_p}\,
\delta_{ij}\delta_{\alpha\beta}
e^{-|t-t'|/\tau_p}.
\end{equation}
This formulation ensures exponentially correlated active forces with variance controlled by the effective temperature $T_{\mathrm{eff}}$. Particle interactions are modeled using the truncated Lennard-Jones potential of the Kob-Andersen binary mixture~\cite{kob1994scaling,kob1995testing}:
\begin{widetext}
\begin{equation}
\phi(r_{ij}) =
\begin{cases}
4\epsilon_{ij}\!\left[\left(\frac{\sigma_{ij}}{r_{ij}}\right)^{12}
-\left(\frac{\sigma_{ij}}{r_{ij}}\right)^6\right], 
& r_{ij} \le r_m, \\[2mm]
\epsilon_{ij}\!\left[A\!\left(\frac{\sigma_{ij}}{r_{ij}}\right)^{12}
- B\!\left(\frac{\sigma_{ij}}{r_{ij}}\right)^6\right]
+ \displaystyle\sum_{p=0}^{3} C_{2p}
\left(\frac{r_{ij}}{\sigma_{ij}}\right)^{2p}, 
& r_m < r_{ij} \le r_c, \\[2mm]
0, & r_{ij} > r_c.
\end{cases}
\label{eq:LJ}
\end{equation}
\end{widetext}
The system consists of $N=10{,}000$ particles in two dimensions with an 80:20 ratio of large to small species at number density $\rho=1.2$ in a periodic box of side length $L=91.287093$. 
Lennard-Jones units are used throughout, with $\varepsilon_{LL}$, $\sigma_{LL}$, and $\tau_0=\sigma_{LL}^2/\varepsilon_{LL}$ serving as the base units. 
The interaction parameters are
\begin{equation}
\varepsilon_{SS}=0.5,\qquad 
\varepsilon_{LS}=1.5,\qquad
\sigma_{SS}=0.88,\qquad 
\sigma_{LS}=0.80.
\end{equation}
The persistence time is varied over $\tau_p \in [2\times 10^{-4}, 1.0]$, spanning the Brownian limit to strongly persistent active dynamics. The effective temperature is varied over $T_{\mathrm{eff}} \in [0.35, 0.40, 0.47, 0.55, 0.65]$.
The equations of motion are integrated using a stochastic velocity-Verlet algorithm~\cite{mannella1989computer} with time step $\Delta t = 10^{-4}$ in MD time unit. 
Each state point is equilibrated for at least $10^6$ steps, followed by production runs of several million steps. To characterize cooperative rearrangements, clusters of highly mobile particles are identified using the DBSCAN algorithm~\cite{ester1996density}, which efficiently detects heterogeneous, non-spherical aggregates without imposing geometric assumptions. For each identified cluster, the gyration tensor is computed and shape descriptors, specifically asphericity and acylindricity-are extracted following Theodorou and Suter~\cite{theodorou1985shape}. This approach provides a robust, geometry-agnostic characterization of cooperative rearranging region (CRR) morphology across different activity regimes.
\section{Structure of CRRs in Active Glasses}
\label{results-1}
\subsection{Identification of excitations and core-shell decomposition}
\label{Transfer function}
\begin{figure}
\includegraphics[scale=0.3]{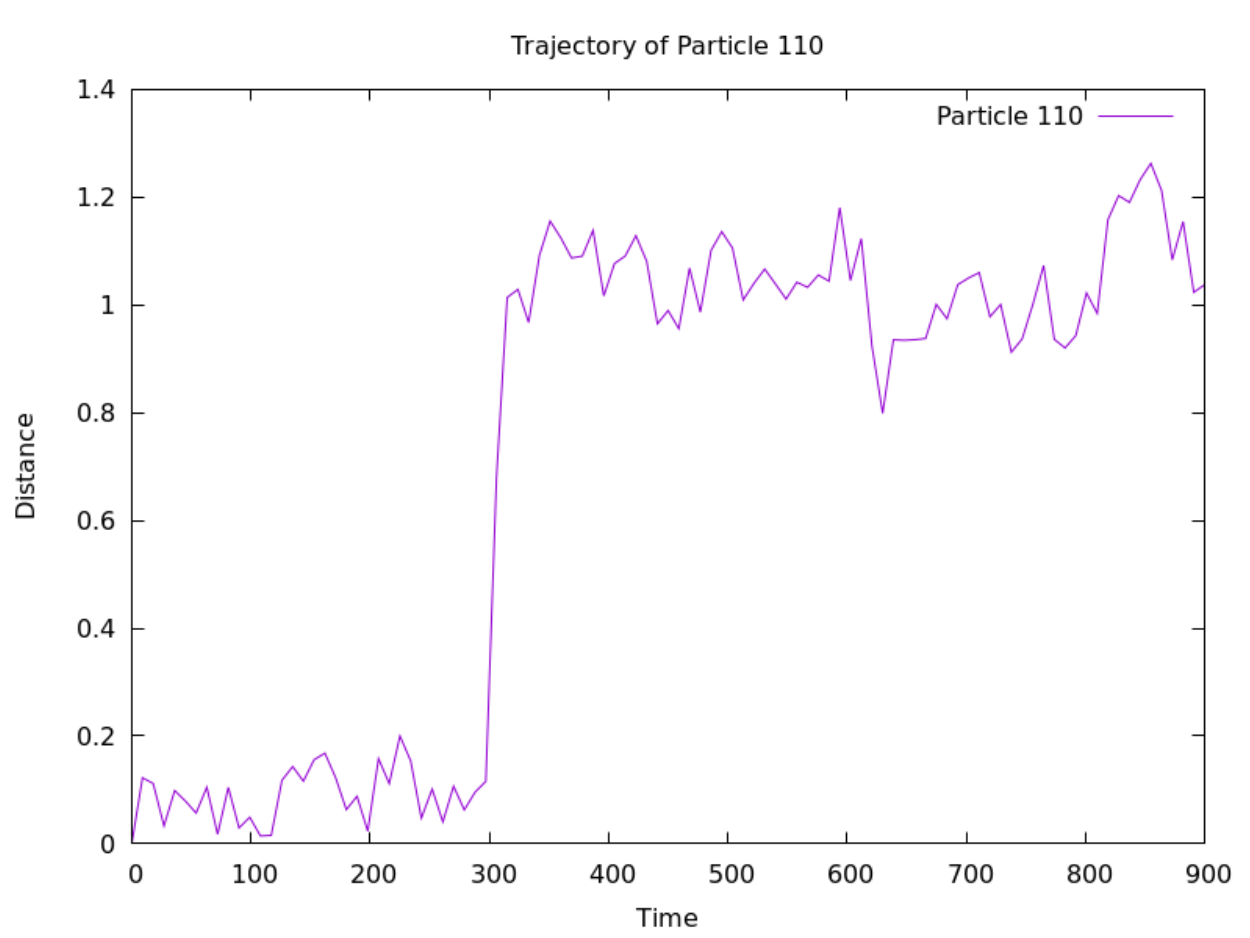}
\caption{Typical particle trajectory associated with an excitation. where time is expressed in MD units.}
\label{fig:pic1}
\end{figure}

\begin{figure}[!h]
    \centering
    \includegraphics[scale=0.9]{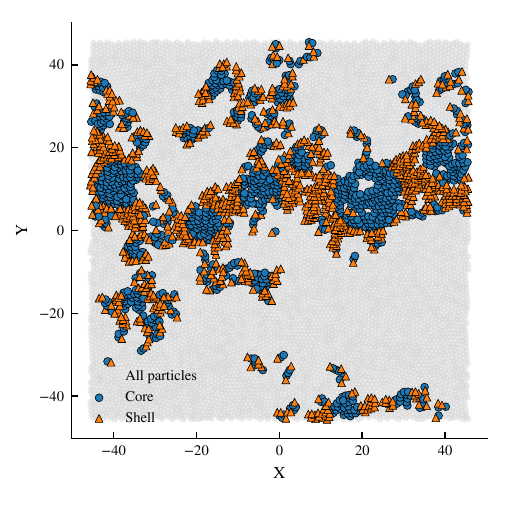}
    \caption{Representative core-shell partitioning of excitation clusters for 
    displacement threshold $a \geq 0.3$ at $T_{\mathrm{eff}} = 0.35$ and
    $\tau_p = 0.1$.}
    \label{fig:core_shell_fraction}
\end{figure}

\begin{figure}
\includegraphics[scale=0.5]{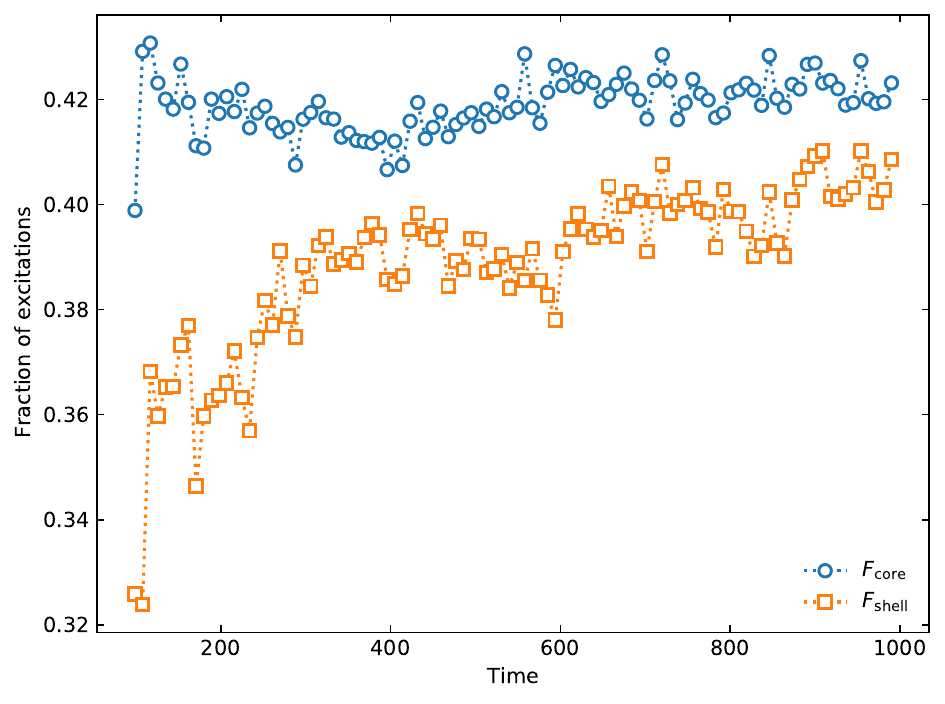}
\caption{Temporal evolution of the fractions of particles in the core and shell regions at $T_{\mathrm{eff}} = 0.35$ and $\tau_p = 0.1$. Time is expressed in MD units.}
\label{fig:pic3}
\end{figure}

Excitations are identified according to Refs.~\citep{gokhale2016localized,gokhale2014growing} by tracking particle displacements over a commitment time $t_a$ and applying a threshold displacement $a$ to isolate persistent rearrangements while excluding transient cage fluctuations. Excited particles are clustered into cooperatively rearranging regions (CRRs) using the density-based clustering algorithm DBSCAN~\citep{ester1996density}, which detects spatially connected mobility events without imposing geometric constraints. The excitation indicator is defined as
\begin{equation}
h_i(t,t_a;a) = 
\prod_{t' = t_a/2 - \Delta t}^{\,t_a/2}
\theta\!\left(\left|\mathbf r_i(t+t') - \mathbf r_i(t-t')\right| - a\right),
\end{equation}
Here, $\theta(x)$ denotes the Heaviside step function, and $\Delta t$ is the microscopic sampling interval. This definition ensures that only sustained mobility events are identified as excitations, as illustrated by a typical excitation trajectory
in Fig.~\ref{fig:pic1}.
Excited particles are grouped into cooperatively rearranging regions (CRRs)
using the density-based clustering algorithm DBSCAN~\citep{ester1996density},
which identifies spatially connected excitations without assuming specific
collective rearrangements. For each CRR, the distance of particle $i$ from the
cluster geometric center $\mathbf R_{\mathrm{cluster}}$ is computed as
\begin{equation}
d_i = \left|\mathbf r_i - \mathbf R_{\mathrm{cluster}}\right|.
\end{equation}
Clusters are partitioned into core and shell regions using the median
distance $d_{\mathrm{median}}$:
\begin{align}
\text{Core: } & d_i \le d_{\mathrm{median}}, \\
\text{Shell: } & d_i > d_{\mathrm{median}}.
\end{align}
This median-based decomposition provides a simple, parameter-free separation between the dense interior of a CRR and its extended periphery.  We have verified that moderate variations in the core-shell partitioning criterion (e.g., using quartile-based thresholds) do not qualitatively alter the observed morphological trends, indicating that the reported behavior is not sensitive to the specific choice of cutoff. Representative snapshots of the resulting core-shell partitioning are shown in Fig.~\ref{fig:core_shell_fraction}, while the temporal evolution of the corresponding core and shell fractions are illustrated in Fig.~\ref{fig:pic3}. These visualizations highlight the spatial organization of
cooperative rearrangements in the active glass.
The decomposition reveals a functional separation within CRRs: the core acts as the excitation center, localizing plastic rearrangements, whereas the shell forms the mechanically responsive layer through which mobility propagates (Section \ref{Shape Analysis}).  As shown below, this structural hierarchy plays a central role in determining how persistence time ($\tau_p$) reorganizes facilitation in competition with $T_{\text{eff}}$. (discussed in Section \ref{Shape Analysis}). Below we focus on displacement distribution ($P(d_{m})$) of CRR particles from the centre of the CRR and the effect of $\tau_p$ on them for various $T_{\text{eff}}$.
\subsection{Modal displacement, polarization, and activity-controlled regimes of cooperative motion}
\label{CRR distribution}
\begin{figure}[!htbp]
\centering
\begin{overpic}[width=0.9\columnwidth]{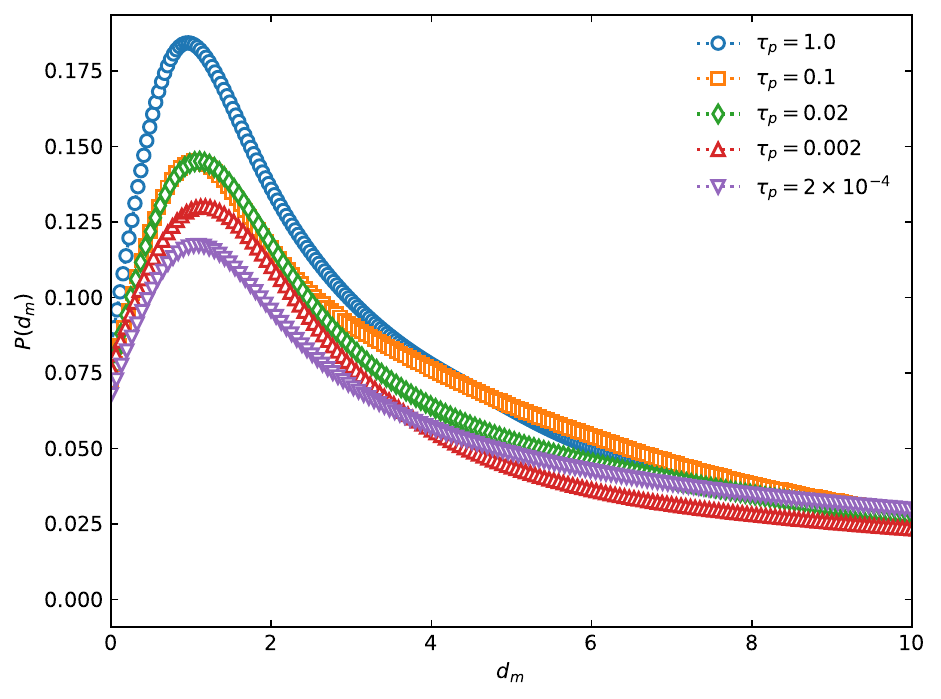}
\put(3,70){\small\textbf{(a)}}
\end{overpic}
\vspace{0.2cm}
\begin{overpic}[width=0.9\columnwidth]{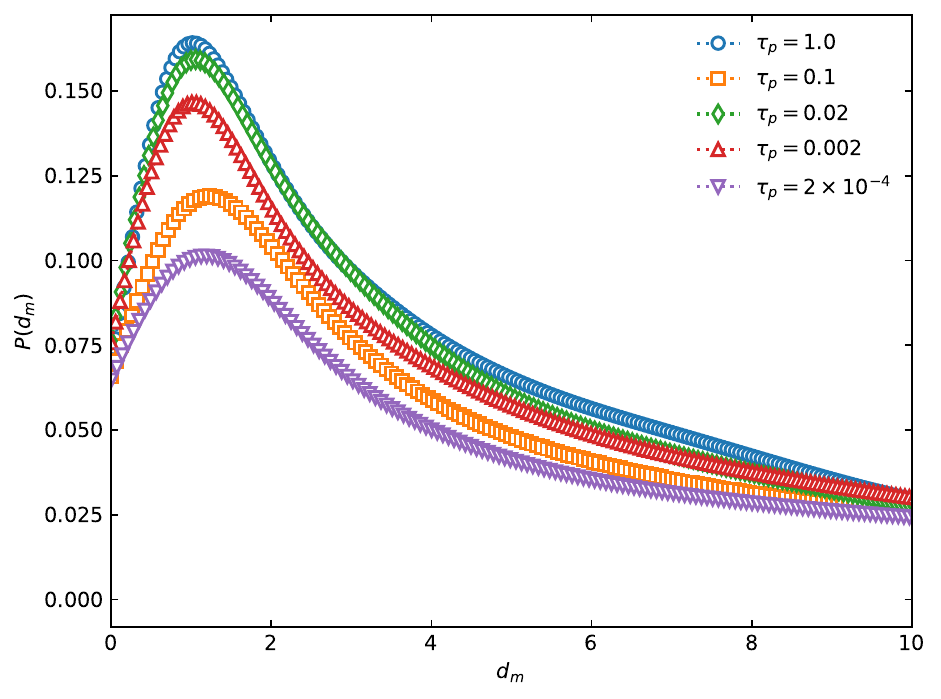}
\put(3,70){\small\textbf{(b)}}
\end{overpic}
\vspace{0.2cm}
\begin{overpic}[width=0.9\columnwidth]{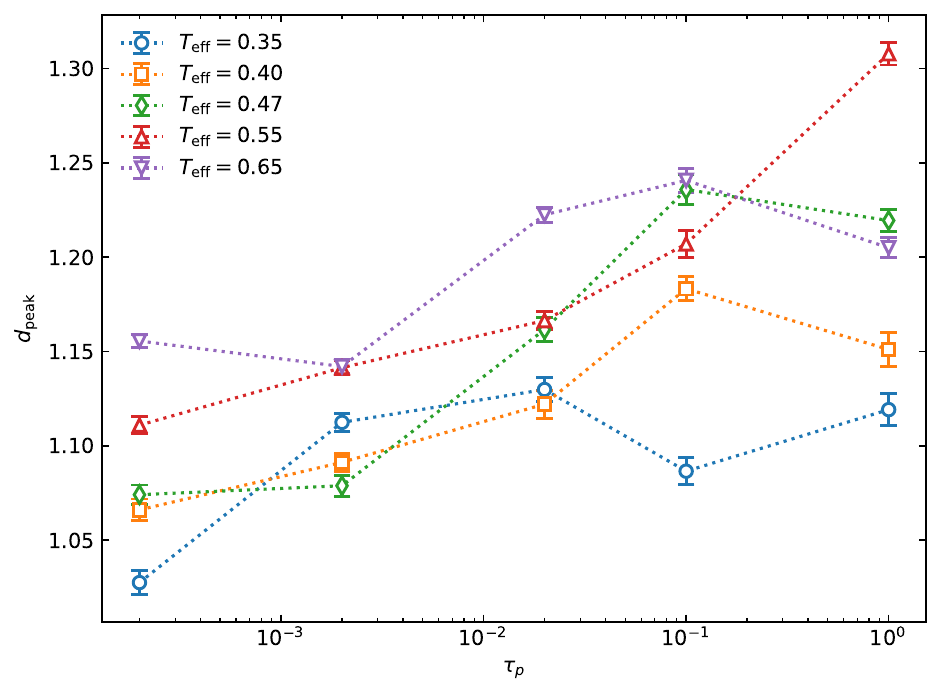}
\put(3,70){\small\textbf{(c)}}
\end{overpic}
\caption{(a, b) Distributions of radial displacements from the CRR center, $P(d_m)$, for different persistence times $\tau_p$ at $T_{\mathrm{eff}}=0.35$ and $0.40$, respectively, normalized such that $\int P(d_m)\,\mathrm{d}d_m = 1$. 
(c) Modal displacement $d_{\mathrm{peak}}$ as a function of $\tau_p$ for several active-noise amplitudes $T_{\mathrm{eff}}$. For lower $T_{\mathrm{eff}}$ (0.35, 0.40, 0.47), $d_{\mathrm{peak}}$ initially increases with $\tau_p$, attains a maximum at intermediate $\tau_p$, and subsequently decreases due to coherent advection at larger $\tau_p$. At $T_{\mathrm{eff}}=0.55$, $d_{\mathrm{peak}}$ increases monotonically, indicating reduced coherent suppression arising from enhanced noise-induced decorrelation relative to persistence-driven coherence. At $T_{\mathrm{eff}}=0.65$, non-monotonic behavior reemerges, with a decrease in $d_{\mathrm{peak}}$ beyond $\tau_p^{*}$, attributable to self-trapped, vortex-like polarized CRRs that suppress relative particle displacements.}
\label{fig:pic4}
\end{figure}

\begin{figure*}[!t]
\centering
\includegraphics[width=0.32\textwidth]{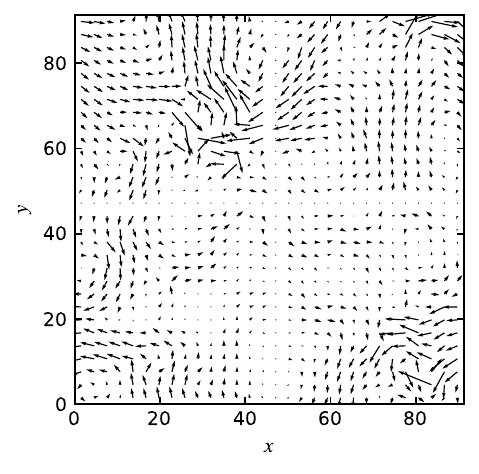}
\includegraphics[width=0.32\textwidth]{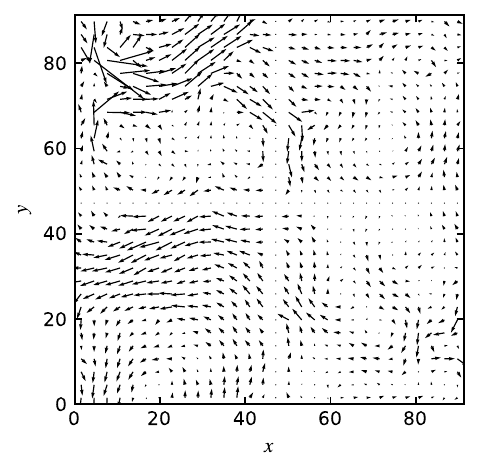}
\includegraphics[width=0.32\textwidth]{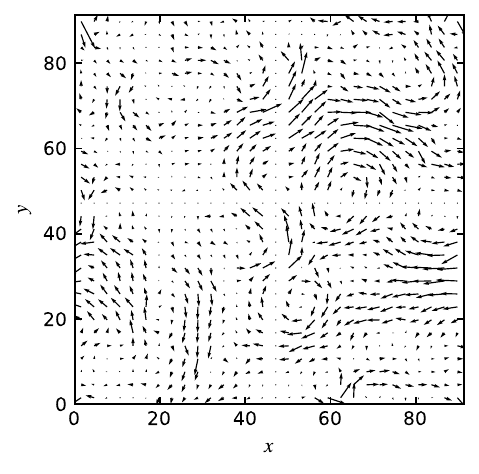}
\caption{Representative activity-controlled regimes of cooperative motion at large persistence time ($\tau_p = 1.0$). Left: at low $T_{\mathrm{eff}} = 0.35$, CRRs exhibit coherent advective motion, wherein persistent alignment drives collective displacement of neighboring particles, suppressing relative motion within the CRR and leading to a reduction in $d_{\mathrm{peak}}$. Center: at intermediate $T_{\mathrm{eff}} \approx 0.55$, enhanced noise fluctuations destabilize coherence-driven advection, preventing the formation of a strongly polarized state. In this regime, persistence primarily promotes relative rearrangements, resulting in a monotonic increase of $d_{\mathrm{peak}}$. Right: at high $T_{\mathrm{eff}} = 0.65$, strong noise fluctuations give rise to dynamically stabilized, vortex-like polarized structures. These self-trapped states suppress relative displacements within CRRs, leading to a decrease in $d_{\mathrm{peak}}$, distinct from the coherence-driven suppression observed at low $T_{\mathrm{eff}}$.}
\label{advec_micro_jam}
\end{figure*}

This study investigates how active driving modifies the internal dynamics of cooperatively rearranging regions (CRRs) by analyzing the distribution of particle distances from the CRR center, denoted by $P(d_m)$. The modal position of this distribution, $d_{\mathrm{peak}}$, corresponds to the most probable radial displacement and therefore characterizes the typical scale of relative rearrangements within a CRR. Because it emphasizes typical rather than extreme displacements, $d_{\mathrm{peak}}$ provides a robust metric for quantifying activity-induced changes in cooperative motion. Fig.~\ref{fig:pic4} (c) illustrates the dependence of $d_{\mathrm{peak}}$ on persistence time $\tau_p$ for various active-noise amplitudes $T_{\mathrm{eff}}$. At small $\tau_p$, $d_{\mathrm{peak}}$ increases with persistence for all noise strengths. In this regime, rapidly fluctuating Ornstein-Uhlenbeck propulsion approximates isotropic agitation. Increasing persistence allows particles to maintain directional bias for longer duration, enabling particles to overcome local cage constraints and thereby enhance cooperative displacements.
At larger persistence times, qualitatively distinct behaviors emerge depending on $T_{\mathrm{eff}}$.\\
\\
At low noise strengths ($T_{\mathrm{eff}} = 0.35, 0.40$), $d_{\mathrm{peak}}$ exhibits a pronounced non-monotonic dependence on $\tau_p$. Following an initial increase, $d_{\mathrm{peak}}$ attains a maximum at an intermediate persistence time $\tau_p^{*}$ and subsequently decreases at larger $\tau_p$. This reduction signals the onset of coherence-dominated dynamics, wherein persistent self-propulsion promotes collective, advective motion of neighboring particles within CRRs. While the overall displacements remain substantial, the relative motion among particles becomes increasingly suppressed, leading to a reduction in $d_{\mathrm{peak}}$. At an intermediate noise strength ($T_{\mathrm{eff}} \approx 0.55$), enhanced stochastic fluctuations weaken the coherence-induced suppression observed at lower $T_{\mathrm{eff}}$. As a result, the system does not develop a strongly polarized, advective state, and $d_{\mathrm{peak}}$ increases monotonically with $\tau_p$. This regime reflects a crossover in which persistence predominantly facilitates relative particle rearrangements, thereby enhancing $d_{\mathrm{peak}}$, without promoting collective, coherence-driven drift of particles. At higher noise amplitude ($T_{\mathrm{eff}} = 0.65$), a non-monotonic dependence of $d_{\mathrm{peak}}$ on $\tau_p$ reemerges, albeit with a distinct underlying mechanism. Strong Ornstein-Uhlenbeck force fluctuations generate transient force-imbalanced configurations that, when combined with persistence, give rise to dynamically stabilized, vortex-like polarized structures. These self-trapped states suppress relative displacements within CRRs, leading to a reduction in $d_{\mathrm{peak}}$ at larger $\tau_p$. The interplay of noise and persistence thus defines distinct activity-controlled dynamical regimes, as summarized schematically in Fig.~\ref{advec_micro_jam}.
Directional coherence within CRRs is quantified using a polarization measure that characterizes collective alignment of particle displacement directions within each excitation cluster. For a CRR containing $N_{\mathrm{CRR}}$ particles, polarization is defined as
\begin{equation}
P_{\mathrm{CRR}} = 
\frac{1}{N_{\mathrm{CRR}}}
\left|
\sum_{i \in \mathrm{CRR}}
\frac{\mathbf v_i}{|\mathbf v_i|}
\right|,
\label{eq:polarization}
\end{equation}
where $\mathbf{v}_i$ denotes the instantaneous velocity of particle $i$ belonging to a given cluster in a CRR. This definition quantifies directional alignment independently of velocity magnitude. Polarization is computed for each excitation cluster and subsequently averaged over clusters and time origins.
\begin{figure*}[!hbtp]
\centering
\includegraphics[scale=0.55]{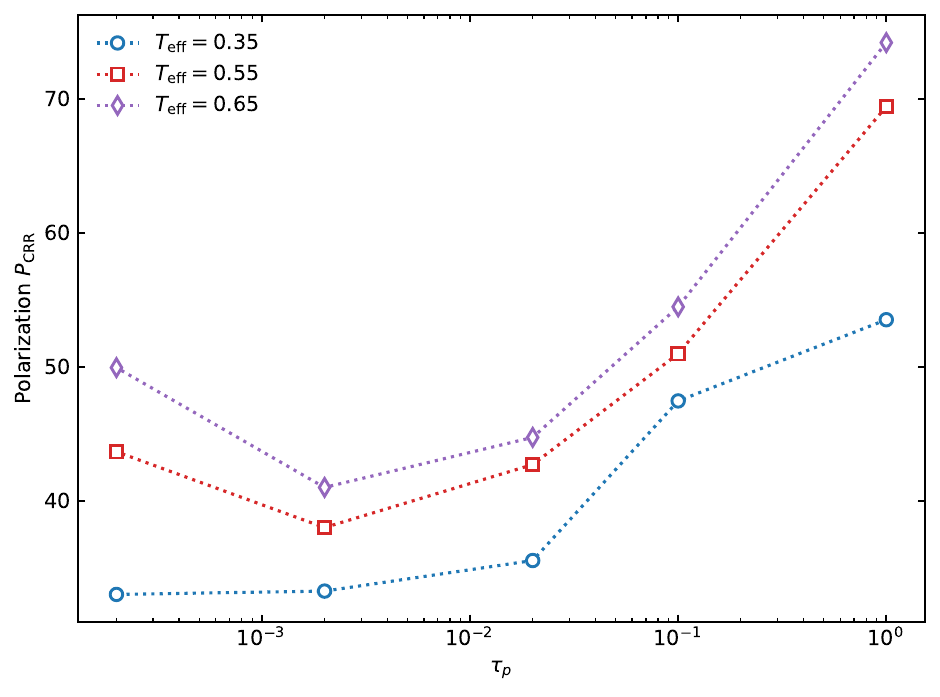}
\caption{Average polarization $\langle P_{\mathrm{CRR}} \rangle$ as a function of persistence time $\tau_p$ for different $T_{\mathrm{eff}}$. Polarization increases with $\tau_p$ for all noise amplitudes, reflecting enhanced directional coherence induced by persistent active forcing.}
\label{polarisation}
\end{figure*}
Fig. ~\ref{polarisation} shows that CRR polarization increases with $\tau_p$ for all $T_{\mathrm{eff}}$, indicating the emergence of directional, coherent motion at large persistence ($\tau_p = 1.0$). Notably, high polarization corresponds to distinct dynamical states depending on the noise amplitude: coherently advecting CRRs at low $T_{\mathrm{eff}} = 0.35$ and self-trapped, vortex-like polarized CRRs at high $T_{\mathrm{eff}} = 0.65$. In contrast, $T_{\mathrm{eff}} \approx 0.55$ serves as a crossover noise strength. In this regime, both coherence-induced suppression and trapping mechanisms are less effective, allowing $d_{\mathrm{peak}}$ to increase with $\tau_p$.

%%%%%%%%%%%%%%%%%%%%%%vorticity inclusion%%%%%%%%%%%%%%%%%%%%%%%%%%%%%%%%%%%%%%%%

\begin{figure}[!hbtp]
\centering
\includegraphics[scale=0.55]{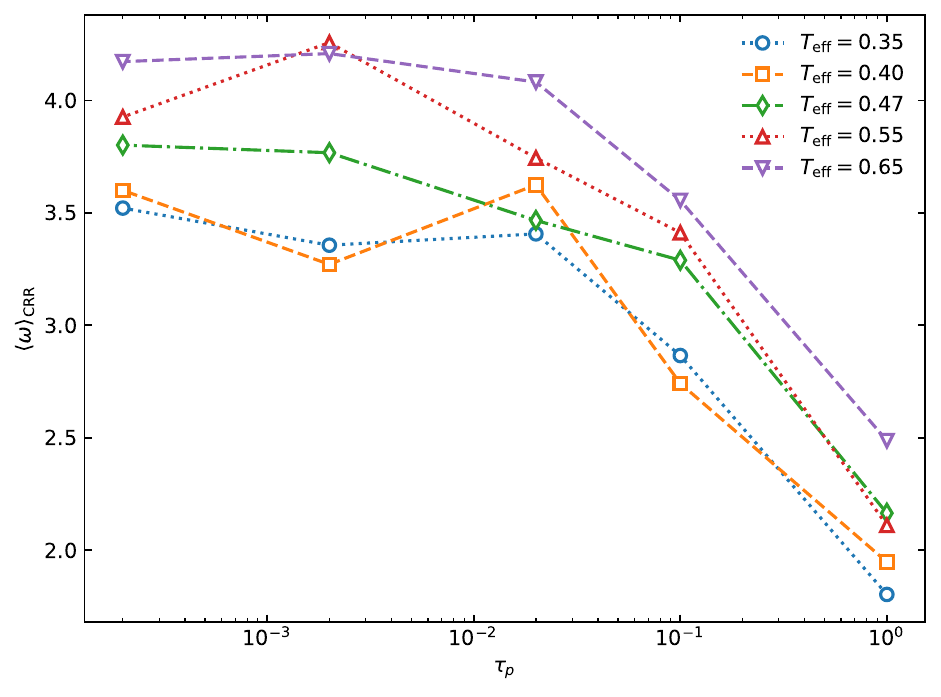}
\caption{
Average vorticity within cooperatively rearranging regions,
$\langle \omega \rangle_{\mathrm{CRR}}$, as a function of persistence time $\tau_p$
for different effective temperatures $T_{\mathrm{eff}}$.
The decrease of vorticity at large $\tau_p$ indicates suppression of internal rotational fluctuations due to coherence-dominated advection at low $T_{\mathrm{eff}}$ and vortex-like trapping at high $T_{\mathrm{eff}}$.
}
\label{vorticity}
\end{figure}

To further characterize the internal dynamics of CRRs, we analyze the average vorticity $\langle \omega \rangle_{\mathrm{CRR}}$ as a function of persistence time $\tau_p$ for different $T_{\mathrm{eff}}$ (Fig.~\ref{vorticity}). The vorticity provides a complementary measure of rotational motion within CRRs, in contrast to polarization which captures directional alignment. We observe that $\langle \omega \rangle_{\mathrm{CRR}}$ decreases with increasing persistence time $\tau_p$ across all effective temperatures. At small $\tau_p$, rapidly decorrelating active forces generate moderate rotational fluctuations within CRRs. As persistence increases, motion becomes increasingly coherent and organized, suppressing internal rotational fluctuations. At low effective temperatures ($T_{\mathrm{eff}} = 0.35, 0.40$), this suppression is associated with coherence-dominated dynamics, where persistent propulsion induces collective advective motion. Although particles move coherently, relative rotational motion within CRRs is reduced, leading to a decrease in $\langle \omega \rangle_{\mathrm{CRR}}$, consistent with the observed reduction in $d_{\mathrm{peak}}$. At high effective temperature ($T_{\mathrm{eff}} = 0.65$), the decrease in vorticity originates from a distinct mechanism. In this regime, large active-force fluctuations combined with persistence generate vortex-like, self-trapped structures. Despite their rotational organization, these structures suppress internal rearrangements due to stabilization by persistent forcing. As a result, the measured vorticity, which reflects fluctuating rotational motion, decreases with increasing $\tau_p$. At intermediate effective temperature ($T_{\mathrm{eff}} \approx 0.55$), corresponding to the crossover regime identified in the $d_{\mathrm{peak}}$ analysis, the vorticity exhibits a weaker dependence on $\tau_p$, reflecting the absence of strong coherence- or trapping-dominated suppression mechanisms. These results demonstrate that both advective and vortex-dominated regimes correspond to dynamically constrained states in which internal rotational fluctuations are suppressed. The simultaneous decrease of $d_{\mathrm{peak}}$ and $\langle \omega \rangle_{\mathrm{CRR}}$ at large $\tau_p$ therefore provides a consistent picture of activity-induced suppression of internal CRR dynamics despite enhanced directional alignment.
The emergence of coherence- and trapping-dominated regimes alters the asymmetry and tail weight of the displacement distribution $P(d_m)$. To quantify these effects, we examine the skewness and kurtosis of $P(d_m)$, which provide complementary information on the statistical structure of cooperative motion and connect naturally to shell occupation probabilities and facilitation lengths discussed in the next section.
%%%%%%%%%%%%%%%%%%%%%%%%%%%%%%%%%%%%%%%%%%%%%%%%%%%%%%%%%%%%%%%%%%
\subsection{Higher-order displacement statistics and dynamical heterogeneity}
\label{kurtosis_and_skewness}
\begin{figure}[h]
\centering
\includegraphics[scale=0.55]{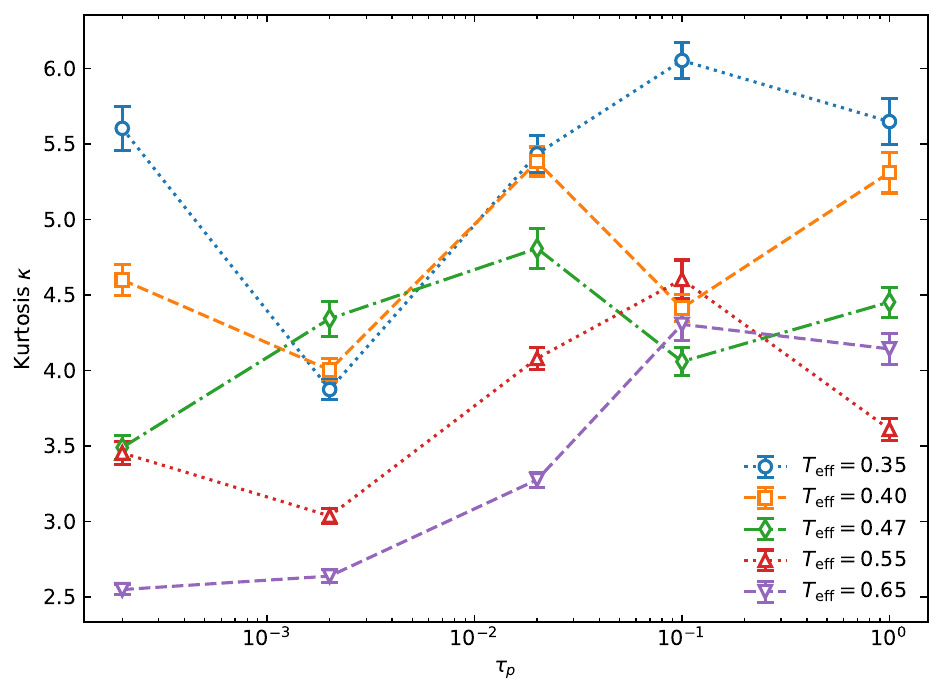}
\caption{Kurtosis $\kappa$ of $P(d_m)$ as a function of persistence time $\tau_p$ for various $T_{\mathrm{eff}}$. The intermediate-$\tau_p$ maximum signals enhanced heavy-tailed statistics associated with rare cooperative rearrangements.}
\label{fig:pic5}
\end{figure}

\begin{figure}[h]
\centering
\includegraphics[scale=0.55]{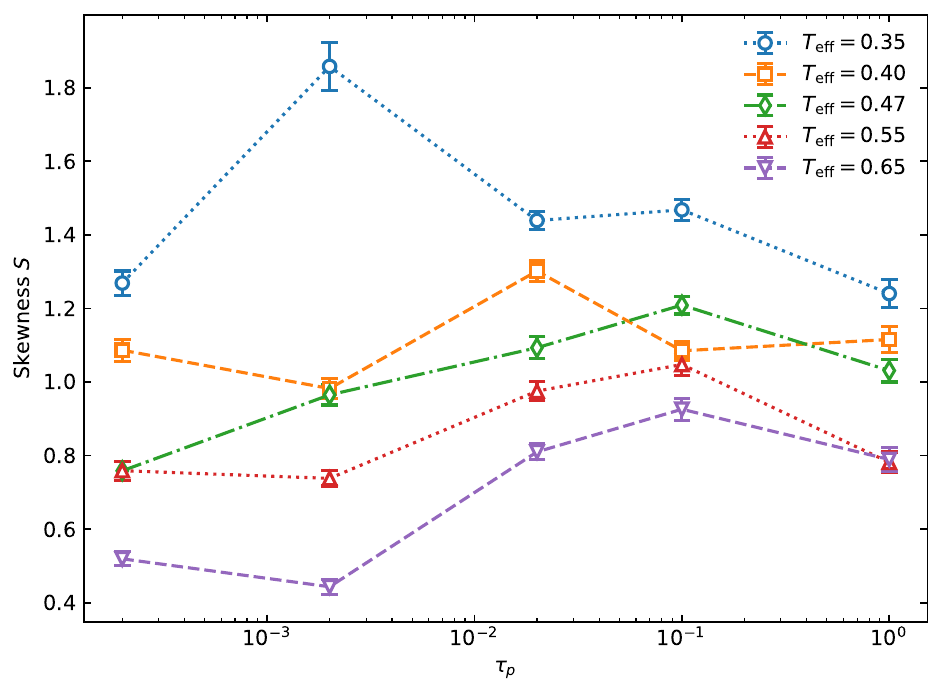}
\caption{Skewness $S$ of $P(d_m)$ versus $\tau_p$. The broad intermediate-$\tau_p$ maximum reflects increasing asymmetry driven by persistent active forcing.}
\label{fig:pic6}
\end{figure}

To quantify activity-induced modifications of cooperative motion beyond typical displacement scales, we analyze higher-order moments of the displacement distribution $P(d_m)$ for various persistence times $\tau_p$ and effective noise amplitudes $T_{\mathrm{eff}}$. In particular, we compute the skewness $S$ and kurtosis $\kappa$, defined as
\begin{equation}
\kappa = 
\frac{\left\langle (d_m - \langle d_m \rangle)^4 \right\rangle}
     {\left\langle (d_m - \langle d_m \rangle)^2 \right\rangle^2},
\qquad
S =
\frac{\left\langle (d_m - \langle d_m \rangle)^3 \right\rangle}
     {\left\langle (d_m - \langle d_m \rangle)^2 \right\rangle^{3/2}},
\label{eq:skew_kurt}
\end{equation}
where $\langle \cdot \rangle$ denotes averaging over particles within a CRR and over independent realizations. Kurtosis quantifies the weight of the distribution tails and thereby probes the prevalence of rare, large rearrangements characteristic of intermittent facilitation, while skewness measures asymmetry and reflects directional bias or advective contributions within CRRs. These quantities complement the modal displacement $d_{\mathrm{peak}}$, polarization, and facilitation length $\xi_{\mathrm{fac}}$ discussed later (Section~\ref{Mobility transfer function}), collectively providing a statistical characterization of cooperative dynamics. Fig. ~\ref{fig:pic5} and Fig. ~\ref{fig:pic6} show $\kappa(\tau_p)$ and $S(\tau_p)$ for several $T_{\mathrm{eff}}$. At small persistence ($\tau_p \lesssim 10^{-3}$), both moments approach their Brownian-limit values. As $\tau_p$ increases into the intermediate regime ($\tau_p \sim 10^{-2} - 10^{-1}$), both $\kappa$ and $S$ increase, signaling the onset of strongly intermittent cooperative motion in which rare, spatially extended rearrangements dominate the displacement statistics. For low effective temperatures ($T_{\mathrm{eff}} = 0.35, 0.40$), $\kappa$ exhibits pronounced non-monotonic behavior: it increases sharply from its Brownian-limit value, attains a maximum at an intermediate persistence time $\tau_p^{*}$, and subsequently decreases. The peak in kurtosis coincides with the persistence window over which $d_{\mathrm{peak}}$, the shell occupation probability $P_{\mathrm{shell}}$ (Section~\ref{Radial_distribution_of_Core_and_Shell}, Fig.~\ref{fig:P_shell_max_core}), and the facilitation length $\xi_{\mathrm{fac}}$ (Section~\ref{Mobility transfer function}, Fig.~\ref{fig:xi_t}) are maximized. This correspondence supports a unified picture in which facilitation is strongest when persistence-induced alignment and noise-driven fluctuations act cooperatively, enhancing rare collective rearrangements. At larger $\tau_p$, $\kappa$ decreases or saturates depending on $T_{\mathrm{eff}}$, indicating suppression of extreme relative displacements as dynamics become coherence- or trapping-dominated. Persistent propulsion increasingly promotes collective advection or stabilized polarized structures, thereby reducing internal rearrangement heterogeneity.
The skewness exhibits related but distinct behavior. At low $T_{\mathrm{eff}}$, $S$ peaks at small-to-intermediate $\tau_p$, reflecting strongly asymmetric displacement distributions arising from advective CRR motion. At higher $T_{\mathrm{eff}}$ ($0.47 \le T_{\mathrm{eff}} \le 0.65$), $S$ increases more gradually with persistence, reaches a broad maximum, and then slightly decreases at large $\tau_p$. This behavior reflects the competition between directional bias induced by persistence and decorrelation arising from strong noise fluctuations.
Importantly, the simultaneous enhancement of skewness and kurtosis at intermediate persistence provides direct statistical evidence for a regime of maximal facilitation. In this regime, displacement distributions become both heavy-tailed (higher $\kappa$) and asymmetric (higher $S$), indicating that cooperative rearrangements are dominated by intermittent, collective events rather than smooth diffusive motion. At the observed largest persistence times ($\tau_p \sim 1$), behavior becomes strongly $T_{\text{eff}}$ dependent. For low $T_{\mathrm{eff}}$ (0.35, 0.40), at this $\tau_p=1.0$,  kurtosis remains elevated, consistent with extended advective CRRs. In contrast, at high $T_{\mathrm{eff}} (0.65)$ for $\tau_p=1.0$, both $\kappa$ and $S$ decrease, reflecting trapping-dominated, vortex-like polarized CRRs in which relative displacements are suppressed despite strong alignment due to large $\tau_p$. Taken together, these results demonstrate that facilitation in active glass formers is intrinsically non-monotonic in persistence time. Maximal facilitation emerges within a finite persistence window where activity ($\tau_p$) is sufficiently long-lived to generate rare collective rearrangements, yet not so persistent as to suppress internal motion through coherent drift or trapping. Higher-order displacement statistics therefore provide a sensitive diagnostic of activity-controlled cooperative dynamics and reveal that persistence time ($\tau_p$) reshapes the full displacement distribution $P(d_{m})$ through its competition with the effective-temperature $T_{\text{eff}}$. Next, we show the effect of the $T_{\text{eff}}$-$\tau_p$ interplay on the core and shell population probabilities to understand this behavior rigorously.
\subsection{Radial distribution of core and shell: $P_{\mathrm{core}}(r)$ and $P_{\mathrm{shell}}(r)$}
\label{Radial_distribution_of_Core_and_Shell}
\begin{figure}[!h]
\centering
\includegraphics[scale=0.52]{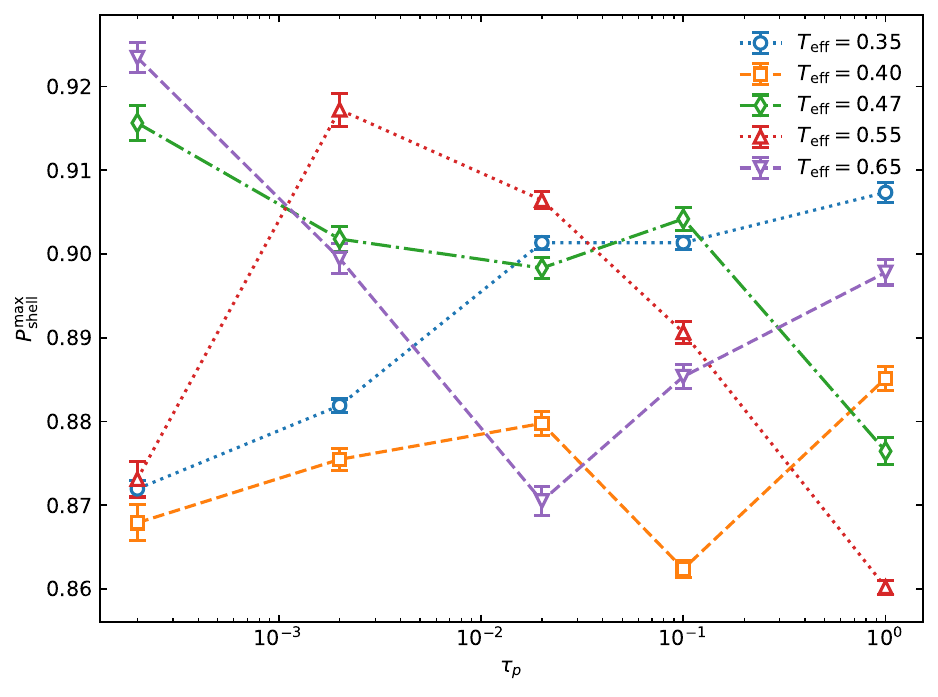}
\caption{Maximum shell occupation probability $P_{\mathrm{shell}}^{\max}$ as a function of persistence time $\tau_p$ for different effective temperatures $T_{\mathrm{eff}}$. The dependence is strongly temperature dependent and non-monotonic, revealing competition between persistence-induced coherence and noise-driven decorrelation.}
\label{fig:P_shell_max_core}
\end{figure}

We analyze the spatial probability distributions of particles in the core ($P_{\mathrm{core}}$) and shell ($P_{\mathrm{shell}}$) regions of cooperatively rearranging regions (CRRs) at fixed effective temperatures $T_{\mathrm{eff}}$ across varying persistence times $\tau_p$. The core and shell are distinguished using a distance-based criterion, where the median distance from the CRR center defines the core radius.
\begin{figure}[!h]
\centering
\includegraphics[scale=0.55]{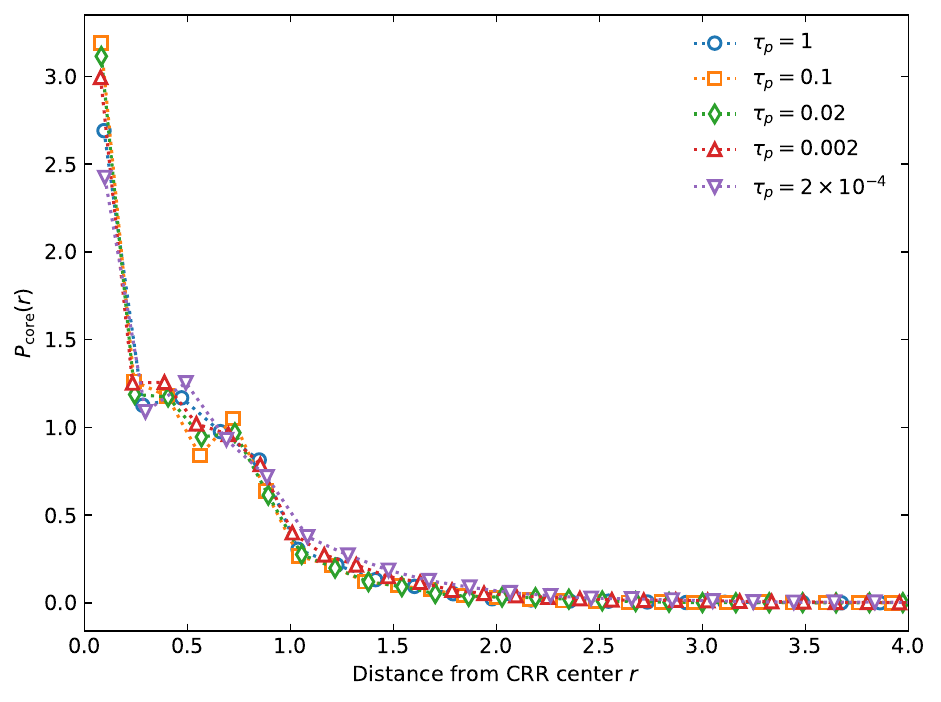}
\caption{Radial distribution $P_{\mathrm{core}}(r)$ from the CRR center at $T_{\mathrm{eff}} = 0.35$ for various $\tau_p$. At fixed temperature, the overall shape of the core distribution remains largely unchanged across different persistence times.}
\label{fig:P_core}
\end{figure}
\begin{figure}[!h]
\centering
\includegraphics[scale=0.55]{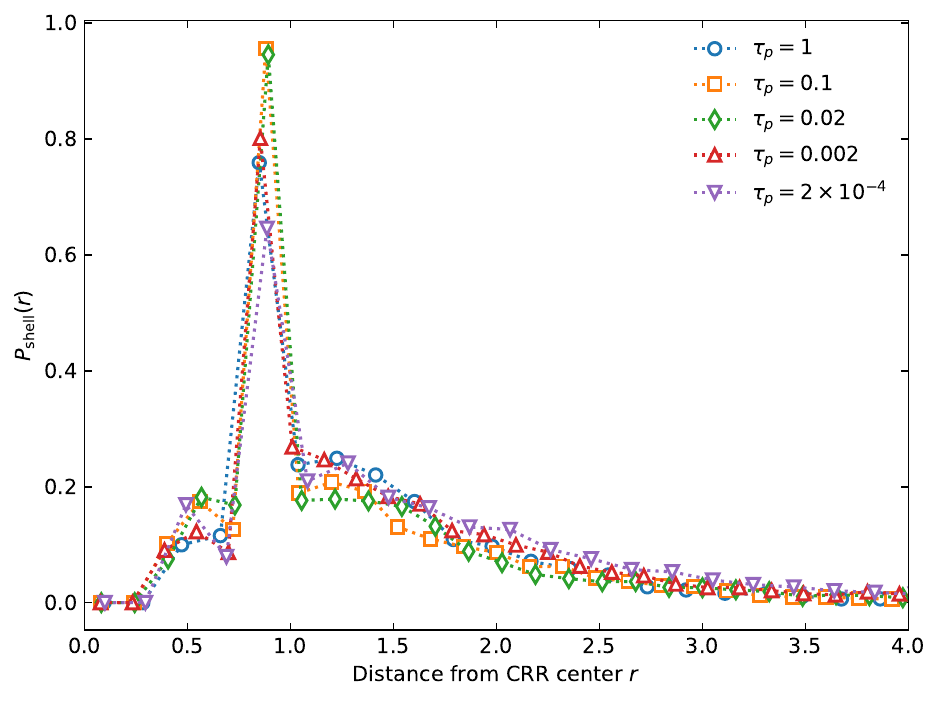}
\caption{Radial distribution $P_{\mathrm{shell}}(r)$ from the CRR center at $T_{\mathrm{eff}} = 0.35$ for various $\tau_p$. While the overall radial extent remains similar, the peak height exhibits a non-monotonic dependence on $\tau_p$.}
\label{fig:P_shell}
\end{figure}
At fixed $T_{\mathrm{eff}}$, the overall shapes of both $P_{\mathrm{core}}(r)$ and $P_{\mathrm{shell}}(r)$ remain nearly invariant with $\tau_p$ (Fig.~\ref{fig:P_core} and Fig.~ \ref{fig:P_shell}), indicating that the geometric partitioning between core and shell is structurally robust against changes in persistence time. However, while the radial size of the core remains stable, the amplitude of $P_{\mathrm{shell}}(r)$ is strongly sensitive to activity (Fig. \ref{fig:P_shell}), reflecting variations in the population of shell-like particles. To quantify this effect, we compute the averaged  maximum shell occupation probability $P_{\mathrm{shell}}^{\max}$ over many realizations
as a function of persistence time $\tau_p$ for different $T_{\mathrm{eff}}$ (Fig. \ref{fig:P_shell_max_core}).
Here $P_{\mathrm{shell}}^{\max}$ serves as a structural proxy for the efficiency of dynamical facilitation. The results demonstrate that the impact of activity ($\tau_p$) on shell formation is strongly $T_{\text{eff}}$ dependent and does not follow a simple monotonic trend. From Fig. \ref{fig:P_shell_max_core}, we see that at low effective temperature ($T_{\mathrm{eff}} = 0.35$), the system is supercooled and particle motion is strongly constrained by cages. In this regime, $P_{\mathrm{shell}}^{\max}$ increases with $\tau_p$, indicating that increasing persistence progressively enhances the population and stability of shell-like structures. Longer-lived active forcing allows particles to push coherently against their cages, strengthening shell-mediated facilitation pathways. At intermediate effective temperatures ($T_{\mathrm{eff}} \approx 0.47$ and $0.55$), $P_{\mathrm{shell}}^{\max}$ exhibits pronounced non-monotonic behavior, with a maximum at intermediate persistence. For small $\tau_p$, activity is too short-lived to stabilize shell structures. At larger $\tau_p$, increasingly coherent advective motion suppresses relative displacements within CRRs, weakening shell formation. At the highest effective temperature ($T_{\mathrm{eff}} = 0.65$) observed here, $P_{\mathrm{shell}}^{\max}$ decreases with increasing $\tau_p$ over a broad range. Here, strong noise ($T_{\text{eff}}$) combined with persistent propulsion  ($\tau_p$) tend to homogenize particle motion, reducing heterogeneity in CRRs and diminishing the distinction between core and shell regions. Overall, the non-monotonic behavior of $P_{\mathrm{shell}}^{\max}$ demonstrates that persistence time $\tau_p$ does not act as a simple control parameter analogous to effective temperature ($T_{\text{eff}}$). Instead, $\tau_p$ couples nontrivially with $T_{\mathrm{eff}}$, selectively enhancing or suppressing shell-mediated facilitation depending on whether persistent activity ($\tau_p$) cooperates with or competes against noise-induced ($T_{\text{eff}}$) decorrelation. Shell occupation therefore provides a sensitive structural indicator of how activity reorganizes facilitation pathways in active glass formers. Next we focus on the facilitation length and the effect of $\tau_p$ on it.
%%%%%%%%%%%%%%%%%%%%%%
\section{Facilitation Length and Mobility Transfer}
\label{Mobility transfer function}
\begin{figure}[H]
\centering
\includegraphics[scale=0.65]{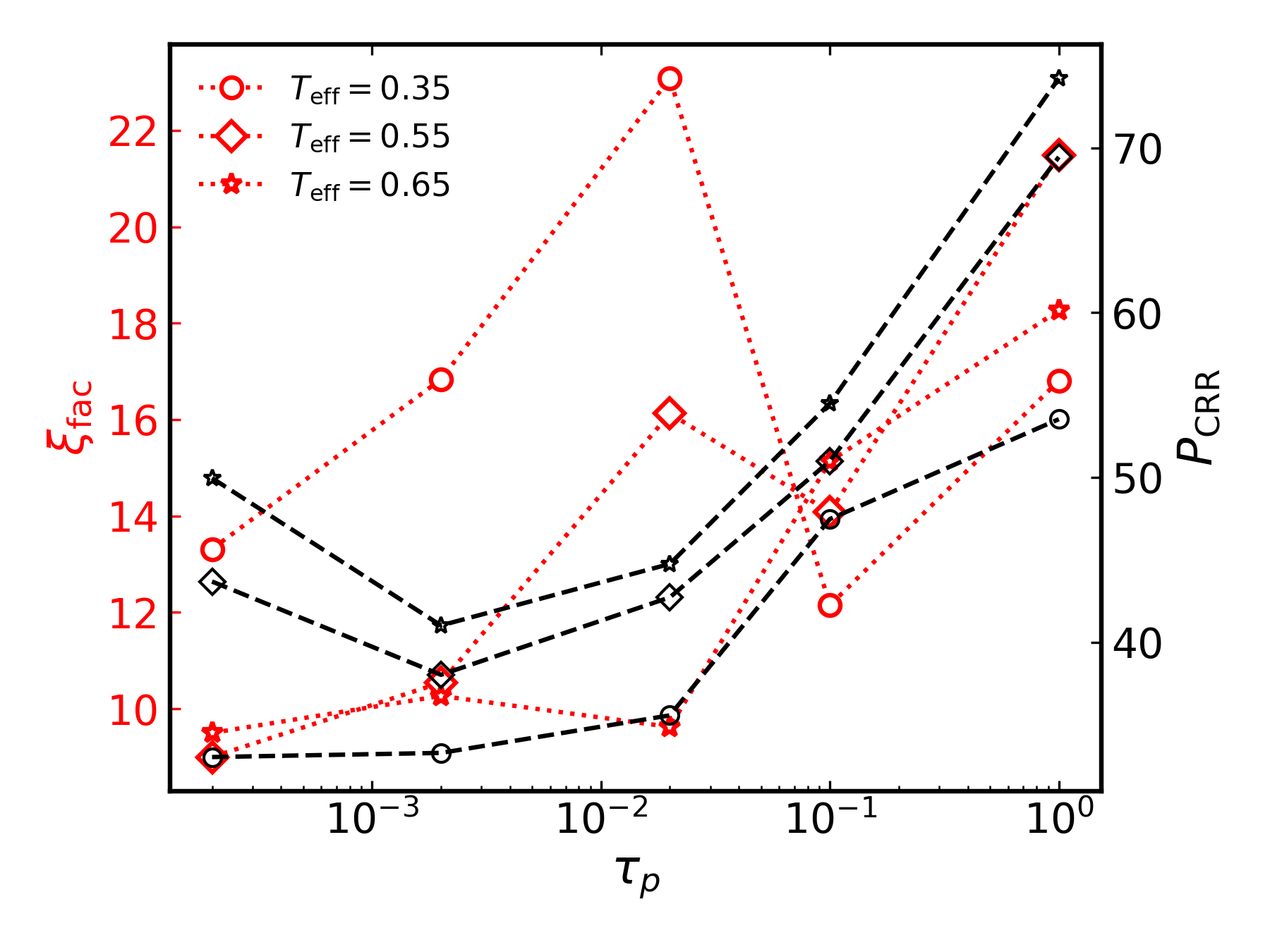}
\caption{Facilitation length $\xi_{\text{fac}}$ (left axis-dotted lines) and CRR polarization $P_{\text{CRR}}$ (right-axis-dashed lines) as a function of persistence time ($\tau_p$) for different effective temperatures. While $\xi_{\text{fac}}$ exhibits a non-monotonic dependence at intermediate $\tau_p$ due to enhancement of relative displacements (elevated $d_{\text{peak}}$, kurtosis, skewness and shell-connectivity), a secondary enhancement is observed at larger $\tau_p \sim 1$. This enhancement coincides with a pronounced increase in polarization, indicating that facilitation transport in this regime is governed by coherence-driven, advective-like dynamics arising from directional alignment (as observed in the reduced $d_{\text{peak}}$, kurtosis and skewness, saturated or lower $P_{\text{shell}}^{\text{max}}$) rather than shell-connectivity within CRRs.}
\label{fig:PCRR-Xi_FAC}
\end{figure}

Facilitation is quantified by measuring the spatial correlation of excitations through the excitation-based mobility transfer function,
\begin{equation}
M(r,\Delta t) =
\frac{
\left\langle 
\mu_i(t+\Delta t)\,\mu_j(t)\,
\delta\!\left(|\mathbf{r}_i-\mathbf{r}_j|-r\right)
\right\rangle
}{
\left\langle \mu_i(t+\Delta t) \right\rangle
\left\langle 
\delta\!\left(|\mathbf{r}_i-\mathbf{r}_j|-r\right)
\right\rangle
}.
\label{eq:mobility_transfer}
\end{equation}
Here, $\mu_i(t)$ denotes the excitation (mobility) indicator of particle $i$ at time $t$, defined as
\begin{equation}
\mu_i(t) =
\begin{cases}
1, & \text{if particle $i$ is mobile in } [t, t+\Delta t], \\
0, & \text{otherwise}.
\end{cases}
\end{equation}
The Dirac delta function selects particle pairs separated by distance $r$, and $\langle \cdot \rangle$ represents an ensemble average over particle pairs and time origins.\\
\begin{figure}[!h]
\centering
\includegraphics[scale=0.5]{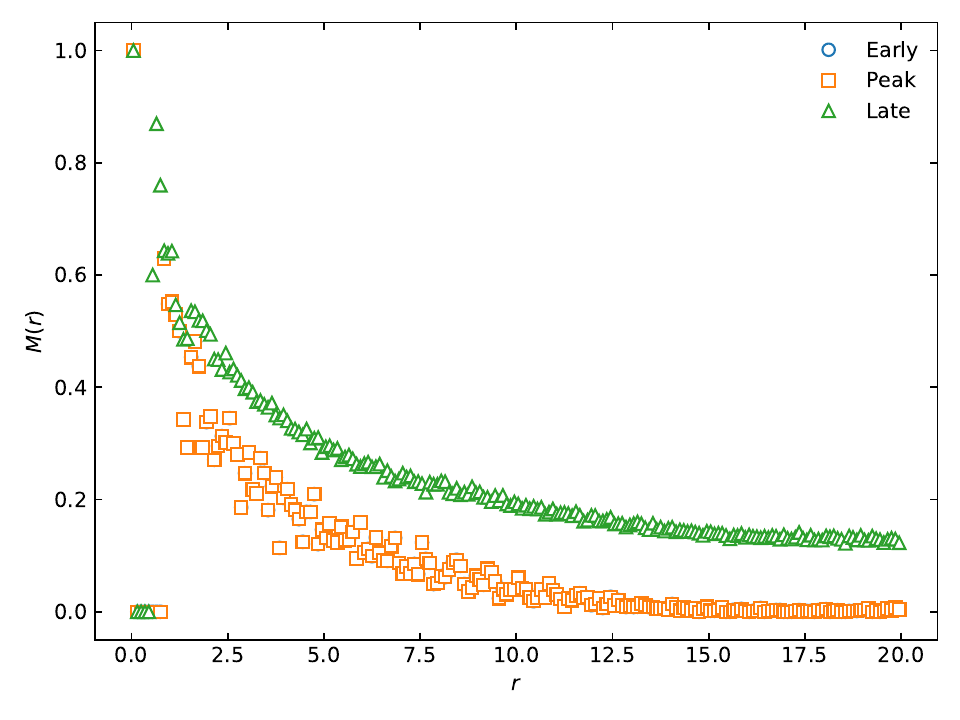}
\caption{Mobility transfer function at early, peak, and late times for $T_{\mathrm{eff}}=0.35$ and $\tau_p=1.0$.}
\label{fig:mobility_trans_at_diff_times}
\end{figure}
\begin{figure}[!h]
\centering
\includegraphics[scale=0.5]{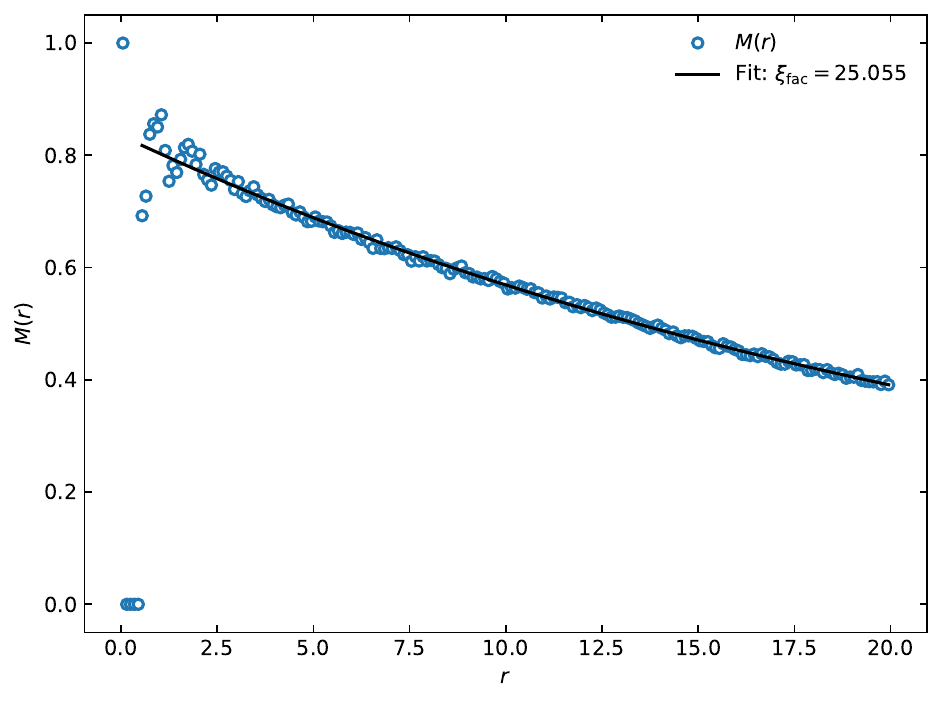}
\caption{Exponential fit $A e^{-r/\xi} + B$ of the mobility transfer function at the peak time for $T_{\mathrm{eff}}=0.35$ and $\tau_p=1.0$.}
\label{fig:fitting_of_mobility_transfer}
\end{figure}
\begin{figure}[!h]
\centering
\includegraphics[scale=0.55]{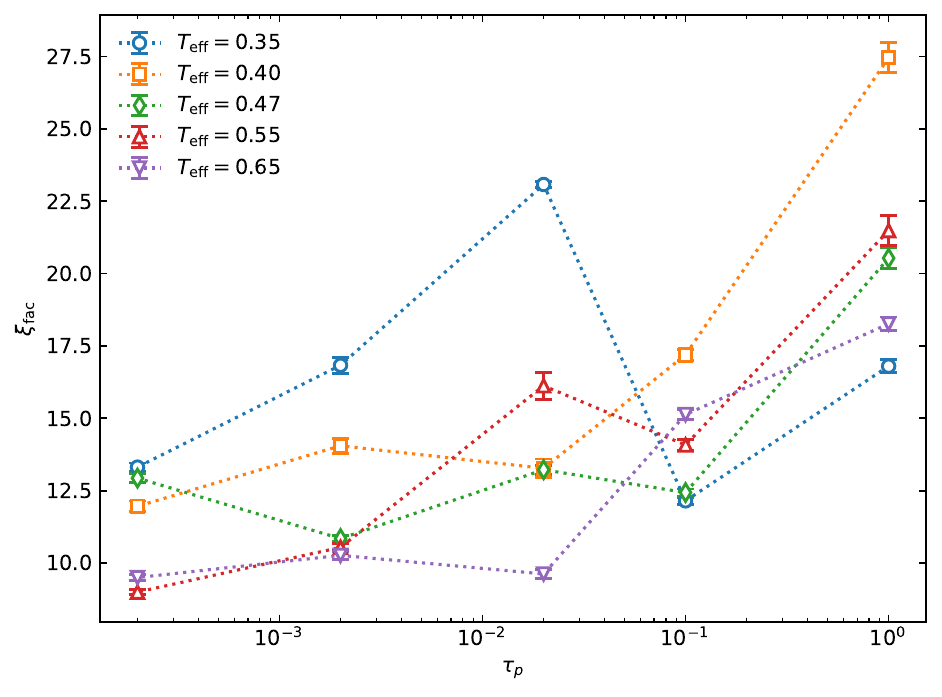}
\caption{Facilitation length $\xi_{\mathrm{fac}}$ as a function of persistence time $\tau_p$ for different effective temperatures $T_{\mathrm{eff}}$. A pronounced non-monotonic dependence on $\tau_p$ is observed at all effective temperatures. A peak at intermediate $\tau_p$ followed by a decrease in $\xi_{\text{fac}}$ is observed at quite higher $\tau_p$. There is also a secondary enhancement of $\xi_{\text{fac}}$ near $\tau_p=1.0$ for all the $T_{\text{eff}}$. The absence of this simple monotonic trend with either control parameter indicates that $\xi_{\mathrm{fac}}$ is governed by the coupled influence of persistence ($\tau_p$) and activity strength ($T_{\text{eff}}$), motivating the introduction of the persistence length $l_p=\sqrt{T_{\text{eff}}\tau_p}$ as a natural microscopic scaling variable.}
\label{fig:xi_t}
\end{figure}
The function $M(r,\Delta t)$ therefore measures the conditional probability that mobility at particle $j$ at time $t$ facilitates mobility at particle $i$ located a distance $r$ away after a time interval $\Delta t$. The facilitation length $\xi_{\mathrm{fac}}$ is obtained from the spatial decay of $M(r,\Delta t)$, thereby linking microscopic excitation dynamics to emergent collective transport. Fig.~\ref{fig:mobility_trans_at_diff_times} shows $M(r,\Delta t)$ at early, peak, and late time dynamics. The peak time $\Delta t^*$ is defined as the time at which facilitation is maximal. For each state point ($T_{\mathrm{eff}}$, $\tau_p$), the mobility transfer function at $\Delta t^*$ is fitted to an exponential form,
\begin{equation}
M(r,\Delta t^*) = A e^{-r/\xi} + B,
\end{equation}
as shown in Fig.~\ref{fig:fitting_of_mobility_transfer}. The extracted decay length $\xi$ at peak time dynamics defines the maximum facilitation length $\xi_{\mathrm{fac}}$, and the maximal value $\xi_{\mathrm{fac}}$ characterizes the spatial extent of excitation propagation. Fig.~\ref{fig:xi_t} presents $\xi_{\mathrm{fac}}$ as a function of persistence time $\tau_p$ for different effective temperatures $T_{\mathrm{eff}}$. A pronounced non-monotonic dependence of $\xi_{\text{fac}}$ on persistence time $\tau_p$ is observed upto $\tau_p \approx 10^{-1}$ across all effective temperatures.
In this regime, the initial increase in $\xi_{\text{fac}}$ at small $\tau_p$ arises from persistence -induced correlations that enhance the propagation of mobility. However, at intermediate persistence times ($\tau_p \approx 10^{-1}$), a reduction in $\xi_{\mathrm{fac}}$ is observed, which can be attributed to the suppression of relative displacements, as reflected in the decrease in $d_{\text{peak}}$. Strongly persistence propulsion induces correlated motion among neighboring particles, leading to coherence-dominated dynamics that reduce the effectiveness of facilitation, since mobility transfer function probes relative rather than absolute motion. As $\tau_p$ increases further toward $\tau_p \sim 1$, a secondary enhancement of $\xi_{\text{fac}}$ is observed. Several mechanisms could, in principle, contribute to this behavior: (i) temporal coupling between  persistence time $\tau_p$ and the structural relaxation time $\tau_{\alpha}$, (ii) coherent advective motion arising from strongly persistent activity, and (iii) enhanced shell connectivity within cooperatively rearranging regions (CRRs). However, direct resonance between $\tau_p$ and $\tau_{\alpha}$ is unlikely to be the dominant mechanism, since $\tau_p$ remains smaller than $\tau_{\alpha}$ throughout the investigated regime. A detailed analysis of structural and dynamical observables further clarifies the origin of this secondary enhancement. While shell occupation probability $P_{\text{shell}}^{\text{max}}$ does not exhibit a corresponding trend of $\xi_{\text{fac}}$ with $\tau_p$, the CRR polarization $P_{\text{CRR}}$ shows (Fig. \ref{fig:PCRR-Xi_FAC}) a strong correlation with the increase in $\xi_{\text{fac}}$. This indicates that shell connectivity alone does not control the enhancement of facilitation as it does in the intermediate $\tau_p$ domain. Instead, the observed behavior at larger $\tau_p \sim 1$ is consistent with a coherence driven mechanism, in which increased directional alignment of particle motion promotes effective transport of mobility. In this regime, strongly persistent activity induces coherent, directed motion within CRRs which can be interpreted as an advective-like transport mechanism. Such coherent motion enables mobility to propagate over larger distances despite the reduction of relative displacements, leading to the observed secondary enhancement of $\xi_{\text{fac}}$. These results demonstrate that, while the initial non-monotonic behavior of $\xi_{\text{fac}}$ is governed by the elevation of relative motion (elevated $d_{\text{peak}}$, kurtosis and skewness) and shell-connectivity (higher $P_{\text{shell}}^{\text{max}}$), the secondary enhancement of $\xi_{\text{fac}}$ is at larger $\tau_p$ arises from dynamical coherence (observed in the reduction of $d_{\text{peak}}$, skewness and kurtosis) rather than static structural connectivity ($P_{\text{shell}}^{\text{max}}$). This highlights the dual role of persistence, which enhances relative displacements and shell-connectivity at intermediate $\tau_p$ while enhancing long-range transport (through the secondary enhancement of $\xi_{}\text{fac}$) at larger $\tau_p \approx 1$.
%%%%%%%%%%%%%%%%%%%%%%%%%%%%%%%%%%%%%%%%%%%%%%%%%%%%%%%%%%%%%% 
\subsection{Scaling of the facilitation length}
\begin{figure}[!h]
\centering
\includegraphics[scale=0.55]{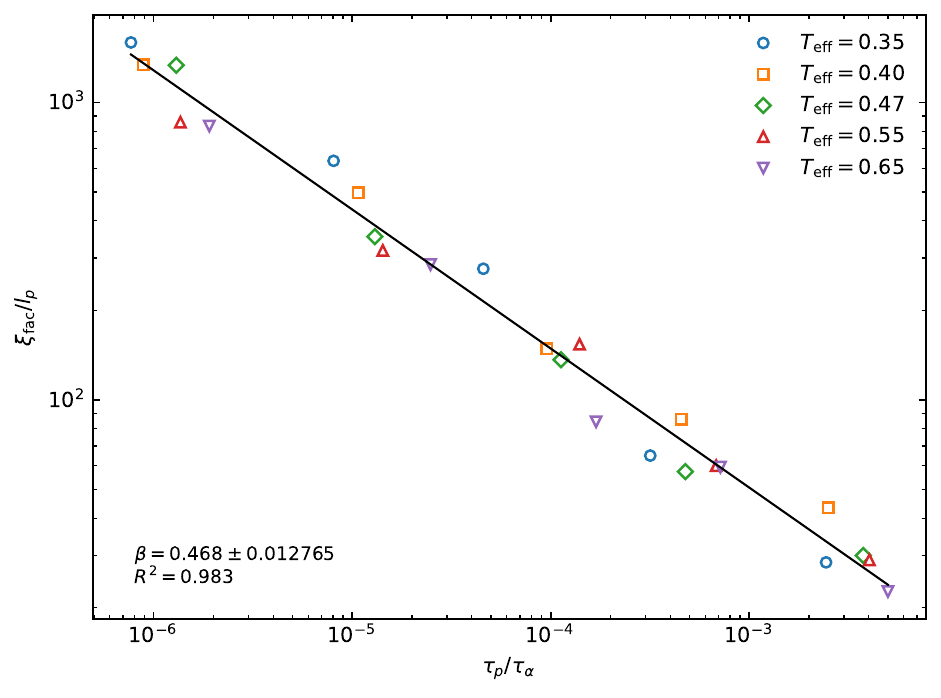}
\caption{Scaling of $\xi_{\mathrm{fac}}/l_p$ versus $\tau_p/\tau_{\alpha}$. 
The collapse across effective temperatures demonstrates a universal 
scaling form governed by the persistence length.}
\label{collapse of xi_lp}
\end{figure}

\begin{figure}[!h]
\centering
\includegraphics[scale=0.55]{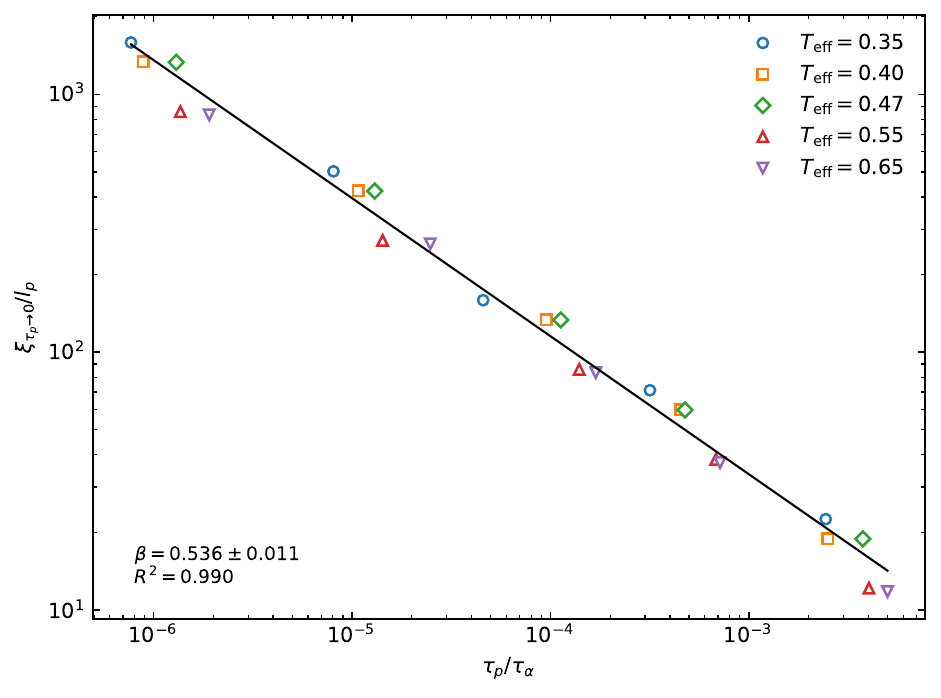}
\caption{Scaling of $\xi_{\mathrm{glass}}/l_p$ versus $\tau_p/\tau_{\alpha}$. 
The similarity of the scaling exponent to that of $\xi_{\mathrm{fac}}$ 
indicates a common underlying dynamical mechanism.}
\label{collapse_xiglass}
\end{figure}

To elucidate how activity controls the spatial transport of dynamical facilitation, we examine the scaling behavior of the facilitation length $\xi_{\mathrm{fac}}$ in the two-dimensional AOUP model. Activity introduces a natural microscopic length scale given by the 
persistence length (in the overdamped case)\\
\begin{equation}
l_p=\sqrt{\frac{k_B T_{\text{eff}} \tau_p}{\gamma}}
\end{equation}
Since, in the reduced dimension $k_B=1$, $\gamma=1$, hence 
\begin{equation}
l_p = \sqrt{T_{\mathrm{eff}} \tau_p},
\end{equation}
which represents the typical displacement accumulated over a persistence time $\tau_p$. This length encapsulates the combined effect of active forcing amplitude ($T_{\mathrm{eff}}$) and memory time ($\tau_p$), making it a natural candidate to rescale spatial transport in the active glass.\\
\\
When $\xi_{\mathrm{fac}}$ is rescaled by $l_p$, data for all effective temperatures collapse onto a single master curve plotted as $\xi_{\mathrm{fac}}/l_p$ versus $\tau_p/\tau_\alpha$ (Fig.~\ref{collapse of xi_lp}), revealing a universal scaling that we discuss below. Here, $\tau_\alpha$ denotes the structural relaxation time. This collapse shows that the combined effects of effective temperature and persistence are fully encoded in the ratio $\tau_p/\tau_\alpha$ once the microscopic length $l_p$ is factored out. Unlike in supercooled liquids, where similar time-length scaling relations hinge mostly on temperature-driven relaxation alone, here the active persistence time fundamentally modifies the universal scaling, highlighting a nonequilibrium route to facilitation that extends beyond passive-glass physics. Such scaling behavior is reminiscent of time-length coupling observed in studies of dynamical heterogeneity in supercooled liquids \citep{berthier2007spontaneous}, while highlighting the distinct 
non-equilibrium origin of facilitation in the present active system.
The collapsed data fit well to a power-law form. The extracted exponent ($0.47$) is close to $1/2$, which is consistent with diffusion-dominated large-scale transport emerging in coarse-grained descriptions of active matter. Such square-root time-length coupling is consistent with coarse-grained descriptions in which persistent forcing renormalizes to effective diffusive transport at long wavelengths \citep{farage2015effective, szamel2014self, fodor2018statistical}. This correspondence is consistent with the theoretical framework of active noise renormalization.
\begin{equation}
\frac{\xi_{\mathrm{fac}}}{l_p}
\sim
\left( \frac{\tau_p}{\tau_\alpha} \right)^{-\beta_{\mathrm{fac}}},
\qquad
\beta_{\mathrm{fac}} \approx 0.47.
\end{equation}
A closely related scaling collapse is observed for the reference length $\xi_{\text{glass}} \equiv \xi_{\text{fac}}(\tau_p \to 0)$, which corresponds to the passive-glass limit where persistence vanishes and activity no longer affects the dynamics. In this regime, $\xi_{\text{glass}}$ provides a clean baseline for comparison, isolating the static facilitation length set purely by glassy relaxation in the absence of active forcing. This clarifies why we refer to this scale as the `glass' reference length as shown in Fig.~\ref{collapse_xiglass}. 
This quantity obeys
\begin{equation}
\frac{\xi_{\mathrm{glass}}}{l_p}
\sim
\left( \frac{\tau_p}{\tau_\alpha} \right)^{-\beta_{\mathrm{glass}}},
\qquad
\beta_{\mathrm{glass}} \approx 0.54.
\end{equation}
The near-symmetric deviation of both exponents around $1/2$ is physically significant. In dynamic critical phenomena, square-root time-length relations arise from Gaussian fixed points characterized by a dynamic exponent $z=2$ \citep{hohenberg1977theory}. Within dynamical facilitation theory, structural relaxation emerges from the diffusion-like spreading of localized mobility excitations at large scale\citep{garrahan2003coarse,chandler2010dynamics,keys2011excitations}. 
Empirically, time-length coupling in supercooled liquids has been established through scaling analyses of dynamical heterogeneity, where growing correlation lengths are found to scale with relaxation time \citep{berthier2005direct}. 
More broadly, coarse-grained descriptions of active matter demonstrate that, despite strong nonequilibrium forcing at microscopic scales, long-wavelength dynamics can renormalize to effective diffusive behavior \citep{marchetti2013hydrodynamics}. 
Although the present results do not constitute a renormalization-group analysis, the observed scaling $\xi_{\mathrm{fac}} \sim \tau_\alpha^{1/2}$ (Eq. \ref{scaling-eq22}) is therefore compatible with diffusion-dominated excitation transport governing structural relaxation in the active glass. Activity reshapes the internal geometry of cooperatively rearranging regions, yet the large-scale time-length coupling remains consistent with diffusive-like transport.
\begin{align}
\xi_{\mathrm{fac}}
&\sim
l_p
\left( \frac{\tau_p}{\tau_\alpha} \right)^{-\beta}
\\
&=
\sqrt{T_{\mathrm{eff}} \tau_p}
\left( \frac{\tau_p}{\tau_\alpha} \right)^{-\beta}.
\label{saling1-eq20}
\end{align}
Rearranging the powers of $\tau_p$ gives
\begin{equation}
\xi_{\mathrm{fac}}
\sim
T_{\mathrm{eff}}^{1/2}
\tau_p^{1/2 - \beta}
\tau_\alpha^{\beta}.
\label{scalin0-eq21}
\end{equation}
For $\beta \approx 1/2$, the factor $\tau_p^{1/2 - \beta}$ is weakly varied in the explored regime. Thus, the dominant scaling reduces to\\
\begin{equation}
\xi_{\mathrm{fac}} \sim \tau_\alpha^{1/2}.
\label{scaling-eq22}
\end{equation}
This square-root dependence reveals an emergent diffusive-like 
time-length relation in the active glass. Unlike hierarchical East-model predictions \citep{sollich2003glassy}, where relaxation grows super-Arrhenius and length-time coupling is logarithmic, the present results exhibit a robust power-law scaling consistent with diffusion-dominated excitation transport. Importantly, this scaling persists despite the pronounced morphological differences and restructuring of 
cooperatively rearranging regions induced by activity. Conceptually, this robustness can be understood by considering that, while activity radically alters the geometry of rearrangements-transforming isotropic, compact clusters typical of passive systems into elongated, polarized structures-the essential transport mechanism remains dominated by stochastic, diffusive-like motion over long timescales. Thus, activity renormalizes the microscopic displacement scale through $l_p$, which remains a genuinely local measure, characteristic of individual particle displacements and is typically of the order of a few particle diameters. By contrast, the facilitation length $\xi_{\mathrm{fac}}$ emerges as a macroscopic collective scale that can exceed $l_p$ by up to two orders of magnitude as observed in our simulations. This explicit separation of scales underscores their distinct roles: the persistence length defines the elementary transport unit, while the structural relaxation time determines the emergent spatial extent of facilitation. The coexistence of strong geometric reorganization with robust diffusive-like scaling suggests that active glasses preserve a passive-like time-length coupling at large scales, even under non-equilibrium forcing. An intriguing open question is whether this diffusive-like scaling survives as the system approaches motility-induced phase separation (MIPS), where additional collective length scales may emerge. Directly probing this behavior in an experiment or simulation would provide a definitive signature to confirm or refute the persistence of the scaling law near MIPS. Exploring that regime may uncover qualitatively new forms of spatiotemporal organization beyond the glassy dynamics characterized here. We find that the observed exponent near $1/2$ indicates that active forcing modifies microscopic displacement scales while leaving the large-scale dynamic exponent unchanged, suggesting robustness of diffusion-controlled facilitation even far from equilibrium. In what follows, we study the morphology variation with $\tau_p$ for various $T_{\text{eff}}$ in the next section.
%%%%%%%%%%%%%%%%%%%%%%%%%%%%%%%%%%%%%%%%%%%%%%%%%%
\section{Morphology of CRRs-Cluster shape metric in two dimension and importance of studying shape metric}
\label{Shape Analysis}
\begin{figure*}[!htbp]
    \centering
    \begin{subfigure}[b]{0.45\textwidth}
        \includegraphics[width=\textwidth]{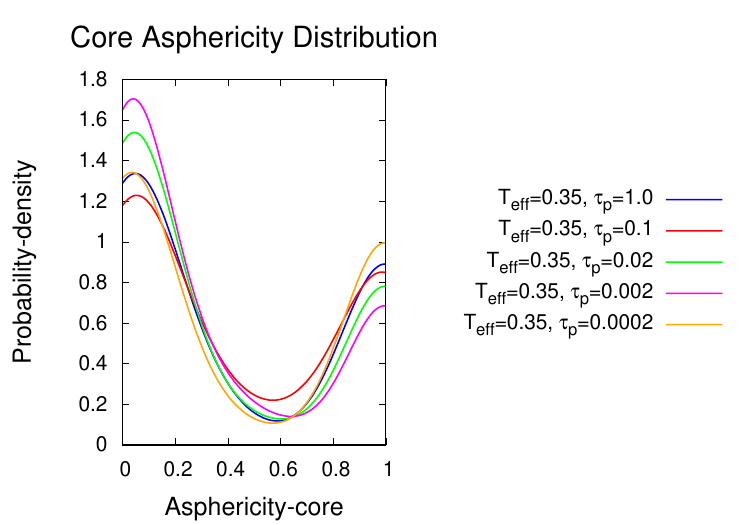}
        \caption{Core Asphericity}
        \label{fig:core_asphericity}
    \end{subfigure}
    \hfill
    \begin{subfigure}[b]{0.45\textwidth}
        \includegraphics[width=\textwidth]{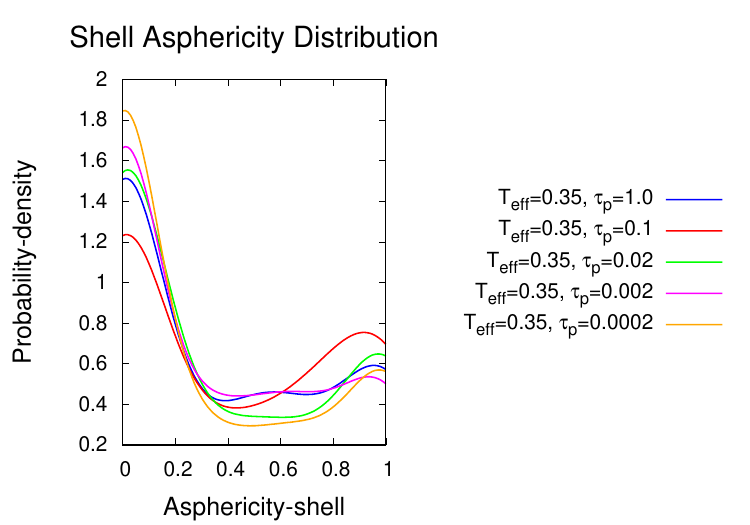}
        \caption{Shell Asphericity}
        \label{fig:shell_asphericity}
    \end{subfigure}
    \vspace{0.5cm} % adjust vertical space between rows
    \begin{subfigure}[b]{0.45\textwidth}
        \includegraphics[width=\textwidth]{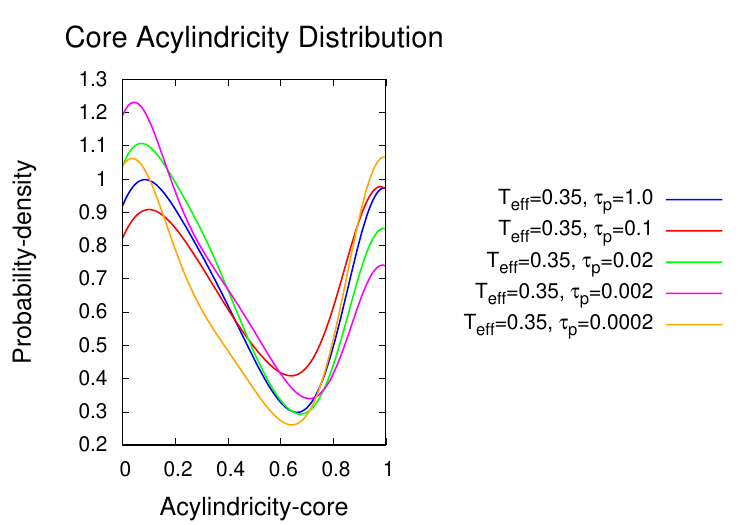}
        \caption{Core Acylindricity}
        \label{fig:core_acylindricity}
    \end{subfigure}
    \hfill
    \begin{subfigure}[b]{0.45\textwidth}
        \includegraphics[width=\textwidth]{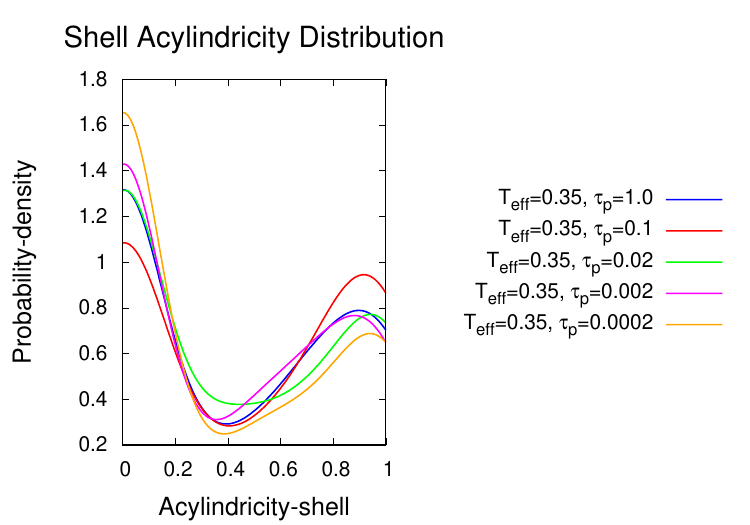}
        \caption{Shell Acylindricity}
        \label{fig:shell_acylindricity}
    \end{subfigure}
    \caption{Morphology distributions (asphericity and acylindricity) of core ((a) and (c)) and shell ((b) and (d)) at $T_{\text{eff}}=0.35$ for various persistence times $\tau_p$. Each panel shows the KDE distributions highlighting the evolution from spherical to elongated morphologies with increasing $\tau_p$.}
    \label{fig:T35_morphology_all}
\end{figure*}

Activity modifies the relative size and structure of cores and shells in CRRs. Core particles correspond to persistent excitations; shell particles are transiently facilitated. While facilitation lengths characterize the spatial extent of cooperative motion, they do not capture the morphology of cooperatively rearranging regions (CRRs). The geometry of these regions whether compact, isotropic, or elongated and string-like plays a crucial role in determining how relaxation events propagate through the system. To address this, we complement our length scale analysis with a shape analysis of CRR cores and their surrounding shells. The distinction between core and shell reflects two regimes of dynamical activity: the core, consisting of the most mobile particles, represents the localized seed of relaxation, whereas the shell captures the cascade of facilitated displacements triggered by the core. By analyzing both, we can disentangle how activity modifies the internal structure of CRRs versus the propagation pathways of facilitation. To quantify CRR morphology, we employ cluster-shape metrics derived from the radius of gyration tensor, including acylindricity and asphericity, which measure deviations from isotropic structure. These metrics allow us to systematically assess whether clusters are compact or elongated, and how such features vary between core and shell regions and across persistence times. This approach provides a statistical characterization of cluster shape that goes beyond average sizes or lengths, offering direct insight into how activity reshapes the geometry of dynamical heterogeneity. The morphology of cooperatively rearranging regions (CRRs) in two dimensions can be quantified using the radius of gyration tensor, defined as
\begin{equation}
S_{\alpha \beta} = \frac{1}{N} \sum_{i=1}^{N} 
\left( r_{i,\alpha} - R_{\alpha} \right)
\left( r_{i,\beta} - R_{\beta} \right),
\qquad \alpha,\beta \in \{x,y\},
\end{equation}
where $r_{i,\alpha}$ denotes the $\alpha$-th Cartesian coordinate of particle $i$, and $R_{\alpha}$ is the corresponding component of the cluster center of mass.  
The tensor $S_{\alpha \beta}$ is real and symmetric, and therefore admits two eigenvalues, 
$\lambda_1^2 \leq \lambda_2^2$, which correspond to the squared principal radii of gyration along the two orthogonal directions. The total squared radius of gyration is then
\begin{equation}
R_g^2 = \lambda_1^2 + \lambda_2^2.
\end{equation}
From these eigenvalues, we define the following shape metrics:
\begin{itemize}
    \item \textbf{Asphericity:}
    \begin{equation}
    b = (\lambda_1 - \lambda_2)^2/(\lambda_1 + \lambda_2)^2,
    \end{equation}
    which measures deviations from circular symmetry. For isotropic clusters $b \to 0$, while elongated clusters yield larger $b$.
    \item \textbf{Acylindricity (2D):}
    In two dimensions, acylindricity is defined by,
    \begin{equation}
    c = |\lambda_1-\lambda_2|/(\lambda_1+\lambda_2),
    \end{equation}
\end{itemize}
In this representation, isotropic cores are characterized by small values of the asphericity $b$, consistent with their compact, nearly spherical geometry. In contrast, the surrounding shells, which mediate facilitation in a more anisotropic manner, exhibit broader distributions shifted toward larger $b$, reflecting the emergence of elongated structures. Complementary to this, the acylindricity $c$, defined from the eigenvalue spectrum of the radius of gyration tensor, quantifies deviations from cylindrical symmetry. Small values of $c$ indicate nearly axisymmetric configurations, whereas larger values correspond to increasingly anisotropic shapes with pronounced planar or branched character. Taken together, the joint statistics of $b$ and $c$ provide a detailed characterization of morphology, revealing a systematic evolution of facilitation pathways from compact, isotropic domains (low $b$, low $c$) to string-like structures (high $b$, low-to-moderate $c$), and ultimately to more complex, anisotropic configurations (high $b$, high $c$) as a function of activity and noise amplitudes.\\
\\
Within dynamical facilitation theory, cooperative relaxation originates from localized excitations that trigger cascades of mobility. In passive systems, these rearrangements are often described as compact or string-like~\citep{garrahan2002glassinessthrough,buhot2002crossover,stevenson2006shapes,hung2019string}, yet their geometric evolution under nonequilibrium driving has not been systematically quantified. To characterize the internal structure of cooperatively rearranging regions (CRRs) in active glass formers, we analyze shape descriptors derived from the gyration tensor, namely asphericity and acylindricity. These quantities quantify anisotropy, enabling a direct mapping between facilitation pathways and cluster geometry. We first examine the morphology variation at $T_{\mathrm{eff}}=0.35$ for different persistence times $\tau_p$ (Fig.~\ref{fig:T35_morphology_all}). Corresponding distributions of the asphericity $b$ and the acylindricity $c$ for other values of $T_{\mathrm{eff}} = 0.40, 0.47, 0.55,$ and $0.65$ are provided in the Supplementary Material.
\\
%%%%%%%%%%%%%%%%%%%%%%%%%%%%%%%%%%%%%%%%%%%%%%%%%%%%%%%%%%%%%%%%
\subsection*{Core and shell morphology (Fig.~\ref{fig:T35_morphology_all}) at $T_{\mathrm{eff}}=0.35$}
\begin{table*}[t]
\caption{Core morphology at $T_{\text{eff}} = 0.35$ for different persistence times $\tau_p$, based on asphericity and acylindricity of KDE distributions.}
\centering
\begin{tabular}{c|p{7.0cm}|p{5.0cm}}
\hline\hline
$\tau_p$ 
& Asphericity and Acylindricity
& Probable Morphology \\
\hline
$0.0002$ 
& Strong peak at very low Acyl ($\sim0.03$-0.05) and Asp ($\sim0.04$); rapidly decaying tail; negligible high-anisotropy population. 
& Spherical / compact; Brownian-like isotropic core. \\
\hline
0.002 
& Very sharp, narrow low-Acyl peak ($\sim0.03$-0.05); Asp peak near $\sim0.04$; extremely weak high-Acyl tail; behavior essentially equilibrium-like. 
& Spherical / compact; minimal activity-induced distortion. \\
\hline
0.02 
& Low-Acyl peak ($\sim0.04$) with a noticeable high-Acyl shoulder (0.7-1.0); Asp still peaked near $\sim0.04$ with small tail. 
& Mostly spherical with intermittent elongations (incipient bimodality). \\
\hline
0.1 
& Dominant high-Acyl peak ($\sim0.95$-0.98); Asp displays substantial high-value weight indicating persistent anisotropic deformation. 
& Predominantly elongated / rod-like; strong activity-stretching effects. \\
\hline
1.0 
& Bimodal Acyl distribution: low-Acyl mode ($\sim0.08$-0.10$)$ and strong high-Acyl mode ($\sim0.9$); Asp shows pronounced high tail. 
& Mixed morphology: coexistence of compact and rod-like conformations at large persistence. \\
\hline\hline
\end{tabular}
\label{table1}
\end{table*}
%%%%%%%%%%%%%%%%%%%%%%%%%%%%%%%%%%%%%%%%%%%%%%%%%%%%%%%%%%%%%%%%%%%%%%%%
The morphology of the core at $T_{\text{eff}} = 0.35$ exhibits a clear and systematic dependence on the persistence time $\tau_p$, as inferred from the asphericity $b$ and acylindricity $c$ distributions. For Brownian dynamics (BD) and the lower persistence values ($\tau_p = 0.0002$ and $0.002$), both $b$ and $c$ display narrow, high-intensity peaks at values close to zero, indicating nearly isotropic and compact cores. The corresponding tails at large anisotropy are comparatively weak, reflecting only rare fluctuations away from spherical symmetry. These regimes are therefore characterized as essentially equilibrium-like and spherical. At slightly larger persistence ($\tau_p = 0.02$), the acylindricity develops a clear high-$c$ shoulder while maintaining a strong low-$c$ peak. This bimodal structure corresponds to predominantly spherical cores with intermittent, activity-driven elongations. The distribution remains centered near the isotropic limit, but the enhanced tail indicates the emergence of transient deformed configurations. At intermediate persistence ($\tau_p=0.1$), the shape distributions become strongly anisotropic-the core acylindricity exhibits a dominant peak at large values $c \approx 1$, and the core asphericity shifts weight toward higher $b$. This behavior signifies sustained deformation of the core into elongated, rod-like structures, consistent with persistent active stresses that prevent relaxation to isotropic condition. At the largest persistence examined ($\tau_p = 1.0$), the core acylindricity becomes bimodal, with a low-$c$ mode near the isotropic regime and a high-$c$ mode near unity. This indicates the coexistence of two distinct morphological families-one consisting of compact, nearly spherical cores, and another of strongly elongated, rod-like configurations. Such coexistence reflects the competition between active persistence, which drives elongation, and the internal relaxation mechanisms that favor isotropic condition. Overall, the core morphology evolves from spherical at very low $\tau_p$ to strongly elongated at intermediate persistence, and finally to a mixed, bimodal regime at the highest persistence studied here in the observed range.\\
\\
At an effective temperature of $T_{\text{eff}} = 0.35$, shell morphology evolves continuously with increasing persistence time $\tau_p$, showing progressively enhanced anisotropic responses. In the Brownian and weakly persistent regimes ($\tau_p = 0.0002, 0.002$), the distributions of shell asphericity and acylindricity are sharply localized near zero. This indicates a compact and isotropic interfacial configuration consistent with equilibrium-like relaxation. At $\tau_p = 0.02$, these distributions broaden modestly, developing finite high-anisotropy tails while maintaining a dominant low-anisotropy peak. This reflects activity-induced surface fluctuations without sustained global distortion. Increasing persistence to $\tau_p = 0.1$ results in substantial weight at larger acylindricity and asphericity values, signaling coherent elongation driven by persistent active stresses. However, the distributions remain broad, indicating ongoing competition between interfacial relaxation mechanisms and active forcing. At the largest persistence ($\tau_p = 1.0$), anisotropy is strongly amplified, yet the shell response remains smoother and less sharply bimodal than that of the interior. This suggests that although activity penetrates the interface, it does not fully destabilize its structural continuity. Overall, the shell undergoes a gradual crossover from isotropic to anisotropic morphology as persistence increases, reflecting its mechanically responsive yet stabilizing character. Future studies could validate these predictions by directly tracking shell shape fluctuations in active-particle experiments or by comparing results with theoretical models incorporating explicit interfacial elasticity, providing further insight into the interplay between activity and interface stability.\\
\\
%%%%%%%%%%%%%%%%%%%%%%%%%%%%%%%%%%%%%%%%%%%%%%%%%%%%%%%%%%%%%%%%%%%%%%%%%%%%%5
\begin{table*}[t]
\caption{Shell morphology at $T_{\text{eff}} = 0.35$ for different persistence times $\tau_p$, based on asphericity and acylindricity of KDE distributions.}
\centering
\begin{tabular}{c|p{7.0cm}|p{5.0cm}}
\hline\hline
$\tau_p$ 
& Asphericity and Acylindricity
& Probable Morphology \\
\hline
BD ($0.0002$) 
& Moderate peak at low Acyl ($\sim0.05-0.08$) and Asp ($\sim0.05$); tail extends slightly to higher values; essentially isotropic. 
& Slightly anisotropic shell; largely spherical. \\
\hline
0.002 
& Sharp low-Acyl peak ($\sim0.04-0.06$); Asp peak near $\sim0.05$; minor high-Acyl tail; nearly equilibrium-like. 
& Nearly spherical shell; minor activity-induced distortions. \\
\hline
0.02 
& Low-Acyl peak ($\sim0.05$) with noticeable high-Acyl shoulder (0.6-0.9); Asp shows a weak tail toward higher values. 
& Spherical to weakly elongated shell; beginning of activity-induced stretching. \\
\hline
0.1 
& Broad Acyl distribution with dominant high-Acyl mode ($\sim0.85-0.95$); Asp has significant high-value weight. 
& Elongated / prolate shell; pronounced activity-induced anisotropy. \\
\hline
1.0 
& Strong bimodal Acyl distribution: low-Acyl mode ($\sim0.08-0.12$) and dominant high-Acyl mode ($\sim0.9-1.0$); Asp shows large tail. 
& Highly anisotropic shell; coexistence of spherical patches and elongated structures. \\
\hline\hline
\end{tabular}
\label{table2}
\end{table*}
%%%%%%%%%%%%%%%%%%%%%%%%%%%%%%%%%%%%%%%%%%%%%%%%%%%%%%%%%%%%%%%%%%%%%%%%%%%%%%%%%%%%%
\begin{figure*}[!htbp]
    \centering

    % ---- Row 1 ----
    \begin{subfigure}{0.48\textwidth}
        \centering
        \includegraphics[width=\textwidth]{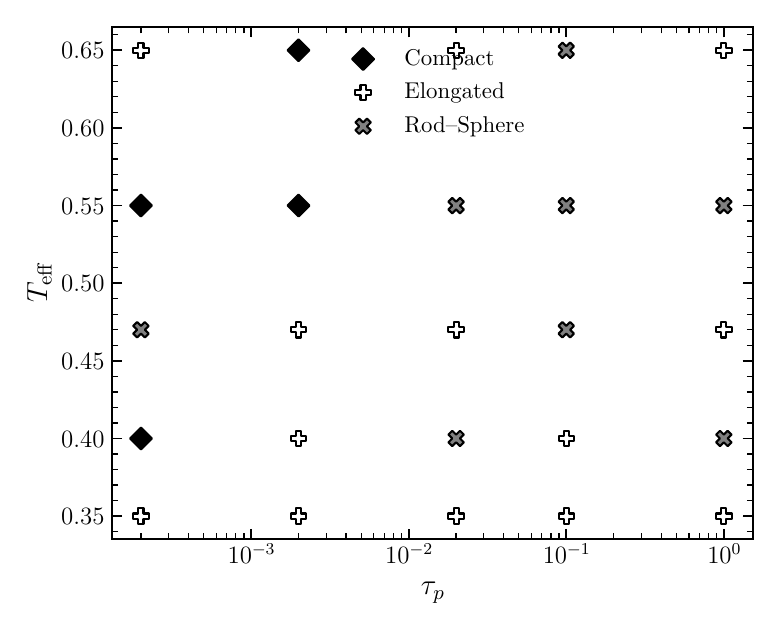}
        
        \caption{Phase diagram of core asphericity across temperature and $\tau_p$. Core asphericity shows a strong, monotonic dependence on $\tau_p$, transitioning from compact shapes at low $\tau_p$ to `rod and sphere' configurations at intermediate $\tau_p$, and fully elongated structures at high  $\tau_p$.}
        \label{fig:a}
    \end{subfigure}
    \hfill
    \begin{subfigure}{0.48\textwidth}
        \centering
        \includegraphics[width=\textwidth]{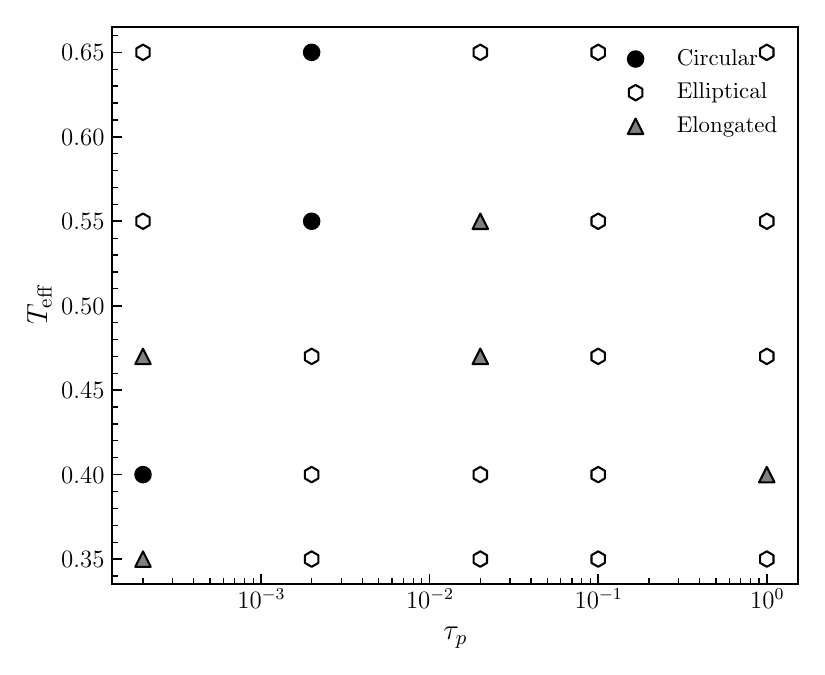}
        \caption{Phase diagram of core acylindricity as a function of temperature and $\tau_p$. Core acylindricity exhibits moderate sensitivity to $\tau_p$, transitioning from circular and elliptical shapes at low $\tau_p$
 to more elongated states at higher $\tau_p$.}
        \label{fig:b}
    \end{subfigure}
    \vspace{0.5cm}
    % ---- Row 2 ----
    \begin{subfigure}{0.48\textwidth}
        \centering
        \includegraphics[width=\textwidth]{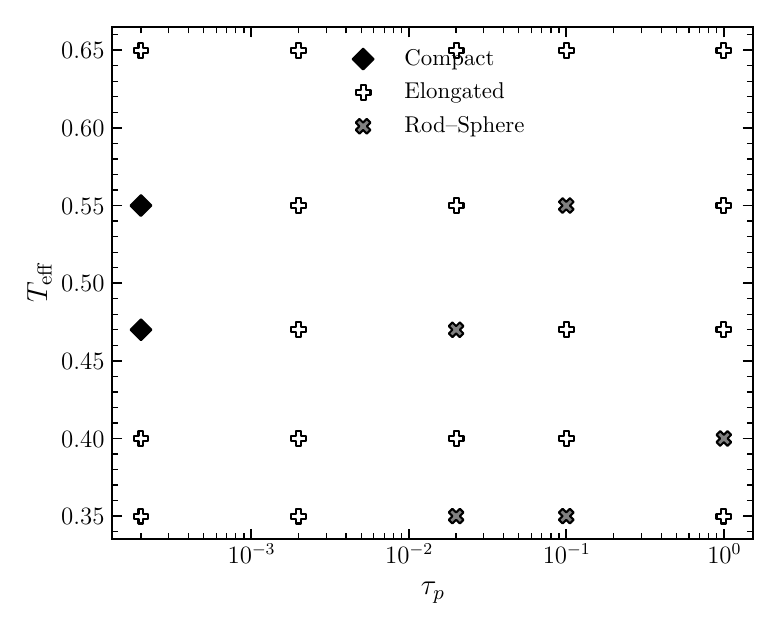}
        \caption{Phase diagram of shell asphericity across temperature and $\tau_p$. Shell asphericity shows relatively weak variation with $\tau_p$, with elongated states dominating most of the phase space. Only low $\tau_p$ and high temperatures yield more compact shell shapes.}
        \label{fig:c}
    \end{subfigure}
    \hfill
    \begin{subfigure}{0.48\textwidth}
        \centering
        \includegraphics[width=\textwidth]{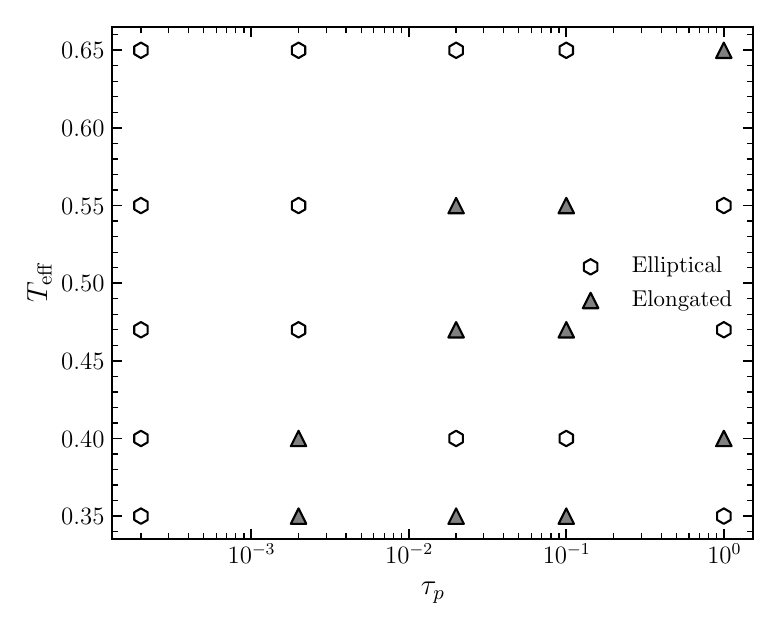}
        \caption{Phase diagram of shell acylindricity as a function of temperature and $\tau_p$. Low $\tau_p$ values produce predominantly elliptical shell shapes, while higher 
$\tau_p$ leads to increasingly elongated morphologies, especially at intermediate temperatures. Shell acylindricity captures the primary activity-driven cylindrical distortion of the shell.}
        \label{fig:d}
    \end{subfigure}
    \caption{Phase diagrams of core and shell based on the percentiles of maxima of Kernel Density Estimation (KDE) distributions of asphericity and acylindricity. The core shows a strong and systematic dependence on $\tau_p$. At low $\tau_p$, it remains compact or circular. Intermediate $\tau_p$ produces mixed `rod and sphere' configurations, indicating the onset of anisotropic condition. At high $\tau_p$, the core becomes fully elongated. Core asphericity captures this progression most clearly, varying much more strongly with $\tau_p$ than acylindricity. The shell exhibits more moderate changes with $\tau_p$. Shell asphericity remains largely unchanged across the parameter space, with elongated shapes dominating. In contrast, shell acylindricity is highly sensitive to $\tau_p$, shifting from predominantly elliptical shapes at low $\tau_p$ to elongated forms at higher $\tau_p$, especially at intermediate temperatures. Thus, $\tau_p$ primarily controls cylindrical stretching in the shell rather than overall deviations from spherical symmetry. Overall, $\tau_p$ drives pronounced shape anisotropy in the core, while shell morphology is governed mainly by activity-dependent cylindrical distortion.}
    \label{fig:four_phase_diagrams}
\end{figure*}
\subsection*{Analysis based on percentile phase diagram}
Now we focus on construction of phase diagram based on the maximum percentiles of the core-shell asphericity-acylindricity distribution. Below we show such plots (Fig. \ref{fig:four_phase_diagrams}). It is envisaged that when $\tau_p$ is small, the active drive fluctuates rapidly and acts like isotropic agitation, causing nearly symmetric stresses on the core. At high effective temperatures, this promotes compact morphologies. At lower effective  temperatures, glassy rigidity prevents complete rounding, allowing weakly irregular shapes to persist. As $\tau_p$ increases, the active force keeps its direction longer, enabling sustained pushing on one side of the core. This directional bias causes anisotropic mass redistribution and more elongated structures, as evidenced by enhanced acylindricity at intermediate $\tau_p$. For larger $\tau_p$, persistent forcing becomes strong and long-lasting enough to cause uneven stress buildup and localized jamming at the core boundary. This leads to smooth rods and irregular, protruded `rod and sphere' morphologies dominating the phase diagrams. Thus, increasing $\tau_p$ does not simply stretch the core. Instead, it transforms activity from isotropic noise into a sustained directional drive that amplifies asymmetry and promotes heterogeneous, arrested shapes, especially at lower effective temperatures where, rather than flowing smoothly into equilibrium, the core freezes mid-stride, arresting its shape in whatever asymmetric pose the persistent activity has imposed.

\subsection{Core-shell morphology interplay and dynamical facilitation}
Our central finding is that dynamic facilitation in these systems is initiated by global deformations of the core and guided by anisotropic responses of the shell, leading to core-driven, shell-modulated excitatory propagation. To quantify the structural characteristics of core and shell particles, we analyzed two shape metrics: asphericity (b), which measures how much a particle's shape deviates from a perfect sphere, and acylindricity (c), which measures deviations from a perfect cylinder. Core particles exhibit $\tau_p$-dependent fluctuations for asphericity, which peaks at intermediate $\tau_p$, indicating elongation along preferred directions; shell particles exhibit $\tau_p$-dependent fluctuations for acylindricity, which shows transitions from elliptical to elongated states, reflecting cooperative reorientation and signaling heterogeneous anisotropic configurations. These variations demonstrate that core morphology is sensitive to the interplay between self-propulsion persistence and noise fluctuations, providing a local structural fingerprint of active dynamics. Analysis of the percentile phase diagrams of core and shell morphology reveals a differential response to variations in the control parameter $\tau_p$. In this context, core's asphericity (see Fig. \ref{fig:a}) fluctuates more with $\tau_p$ compared to shell's asphericity (see Fig. \ref{fig:c}), whereas shell's acylindricity (see Fig. \ref{fig:d}) fluctuates more with $\tau_p$ than that of core (see Fig. \ref{fig:b}). Thus, specifically, the core exhibits larger fluctuations in asphericity, indicating enhanced mechanical softness along its principal axes, which enables it to undergo substantial global shape changes. In contrast, the shell demonstrates greater fluctuations in acylindricity. This reflects flexibility in lateral or cross-sectional modes that allows the shell to accommodate core-driven deformations while maintaining overall axial stability. This hierarchical interplay suggests that morphological transitions are initiated by the core and subsequently mediated by the shell, thereby facilitating dynamical propagation of shape changes in a coordinated and energetically efficient manner. Thus, we conclude that the shell efficiently transmits excitations while preserving its morphology, acting as a mechanical scaffold and spatially extended facilitator of cooperative dynamics. The combined analysis of asphericity and acylindricity thus demonstrates that, in active glass formers, cooperative rearrangements emerge from the dynamics of excitation propagation rather than from shape transformations, extending the DF framework to non-equilibrium active systems. Furthermore, the optimal facilitation observed at intermediate $\tau_p$, where $P_{\text{shell}}$ and $\xi_{\text{fac}}$ peak, quantitatively links persistence time and noise amplitude to the efficiency of mobility propagation, revealing a precise timescale for maximal cooperative activity. In other words, the core deforms globally, and the shell deforms anisotropically, so dynamic facilitation becomes core-driven and shell-modulated, with the core enabling widespread mobility while the shell channels it in a directional manner. Hence, facilitation events originate in the core, penetrate outward, and then propagate directionally within the shell.
%%%%%%%%%%%%%%%%%%%%%%%%%%%%%%%%%%%%%%%%%%%%%%%%%%%%%%%%%%%%%%%%%%%%%%%%%%%%%%%%%%%%%
\section{Results and Conclusions}
\label{conclu}
The morphology of core and shell particles was analyzed across a range of persistence times ($\tau_p$) and effective temperatures ($T_{\text{eff}}$) using two shape metrics: asphericity (Asp) and acylindricity (Acyl). Peaks in the respective distributions were classified into distinct phases using dynamic percentile thresholds. Phase diagrams were constructed separately for the core and shell to investigate correlations between shape fluctuations and particle dynamics. The core shows a clear, monotonic evolution with increasing $ \tau_p$. At low $ \tau_p$, it remains compact or nearly circular, indicating minimal internal deformation. In the intermediate regime, the core often adopts mixed `rod-and-sphere' configurations, signaling the emergence of internal anisotropy driven by persistent active stresses. At high $ \tau_p$, the core becomes increasingly elongated. This progression is best captured by core asphericity, which depends more strongly on $\tau_p$ than acylindricity. Overall, core morphology evolves from compact to mixed anisotropic and finally to fully elongated structures as persistence increases. In contrast, the shell shows moderate morphological variation with $ \tau_p$. Shell asphericity remains stable across most of the parameter space, with elongated configurations dominating a wide range of $T_{ \text{eff}}$ and $\tau_p$. However, shell acylindricity is highly sensitive to persistence. At low $ \tau_p$, the shell is mostly elliptical, while at higher $ \tau_p$ it shifts toward more elongated states, especially at intermediate $T_{ \text{eff}}$. These observations indicate that $ \tau_p$ mainly controls cylindrical stretching within the shell while preserving its overall anisotropic character. Effective temperature acts as a secondary modulator, shifting the balance between elliptical and elongated regimes without qualitatively altering the shell structure. These results also show that core morphology is strongly governed by $ \tau_p$-dependent shape anisotropy, while shell morphology is more subtly regulated by activity-induced distortion. The contrasting sensitivities of core and shell illustrate how persistence reorganizes internal plastic regions and surrounding transport pathways in distinct but coupled ways. Importantly, the regimes of maximal core anisotropy and enhanced shell connectivity coincide with the peak in facilitation length identified in Section \ref{Mobility transfer function}, indicating that morphology and mobility propagation are tightly linked. In this framework, the core functions as the primary locus of structural rearrangement, while the shell provides spatially extended pathways that mediate mobility transfer. Persistence, therefore, reshapes cooperative dynamics not only by modifying excitation density but also by reorganizing the internal geometry of CRRs. Importantly, the correlations between facilitation length, polarization, higher-order displacement moments, and core-shell morphology are systematic across five effective temperatures and nearly four decades of persistence time. The observed trends are consistent across all effective temperatures and persistence times explored in this study, indicating that the reported activity-induced restructuring of facilitation is a robust nonequilibrium effect rather than a parameter-specific anomaly. Thus in a two-dimensional athermal Ornstein-Uhlenbeck active glass former, activity reorganizes facilitation at the level of excitation morphology without altering its large-scale transport law. Shell occupation probability, polarization, and higher-order displacement moments reveal a pronounced non-monotonic dependence on persistence time $ \tau_p$. We identify a regime of maximal cooperative transport at intermediate persistence. However, upon rescaling by the persistence length $l_p=\sqrt{T_{\text{eff}}\tau_p}$, $\xi_{\text{fac}}$ collapses onto a universal master curve described by $\xi_{\mathrm{fac}} \sim l_p \left(\frac{\tau_p}{\tau_\alpha}\right)^{-\beta}$,$ \beta \approx 1/2$. This relation holds throughout the regime where $\tau_p < \tau_\alpha$. It implies an emergent diffusive-like time-length coupling $\xi_{\mathrm{fac}} \sim \tau_\alpha^{1/2}$ despite pronounced activity-induced morphological reorganization. The results demonstrate that active forcing reshapes the geometry and connectivity of facilitation pathways while preserving the underlying diffusive excitation transport and identify the persistence length ($l_p$) as the key microscopic control parameter for cooperative relaxation, indicating that activity renormalizes microscopic structure without altering the large-scale dynamic universality class.
\begin{acknowledgments}
I express my gratitude to professor Manoj Gopalakrishnan and professor V. Vishwas for their valuable feedback and suggestions. All simulations were conducted on the AQUA cluster at IIT Madras. I also thank my friend Anjali M and professor Debashis Chatterjee for reviewing the manuscript and providing additional suggestions.
\end{acknowledgments}
\bibliographystyle{apsrev4-2}
\bibliography{refs}
\end{document}